\documentclass[twocolumn]{autart}   
\usepackage{graphicx}          
\usepackage{array}
\usepackage{amsmath,amssymb,amsfonts}
\usepackage{graphicx}
\usepackage{hyperref}
\usepackage{textcomp}
\usepackage{url}           
\usepackage{nicefrac}      
\usepackage{algorithm}
\usepackage[authoryear]{natbib}
\usepackage{bm}
\usepackage{multirow} 
\usepackage{algpseudocode}
\usepackage{longtable}
\usepackage{textcomp}
\usepackage{longtable} 
\usepackage{supertabular} 
\usepackage{subcaption}
\usepackage{tabularx}
\usepackage{comment}
\usepackage{mathrsfs} 
\usepackage{float}
\usepackage{booktabs} 
\usepackage{makecell} 
\usepackage{tikz}
\usepackage{wrapfig}
\usetikzlibrary{arrows.meta, positioning, calc}
\newtheorem{definition}{Definition}
\newtheorem{theorem}{Theorem}

\newtheorem{corollary}{Corollary}

\newtheorem{assumption}{Assumption}
\newtheorem{remark}{Remark}

\allowdisplaybreaks[4]
\makeatletter
\providecommand{\qedsymbol}{\ensuremath{\square}}
\@ifundefined{proof}{
  \newenvironment{proof}[1][Proof]{
    \par\noindent\textit{#1}. \normalfont
  }{
    \hfill\qedsymbol\par
  }
}{}
\makeatother

\begin{document}

\begin{frontmatter}

\title{Scalable Synthesis and Verification of String Stable Neural Certificates for Interconnected Systems\thanksref{footnoteinfo}} 

\thanks[footnoteinfo]{
Corresponding author Kaidi Yang.}

\author[CEE]{Jingyuan Zhou}\ead{jingyuanzhou@u.nus.edu},  
\author[Amherst]{Haoze Wu}\ead{hwu@amherst.edu},  
\author[IORA]{Haokun Yu}\ead{yuhaokun@u.nus.edu}, 
\author[CEE]{Kaidi Yang}\ead{kaidi.yang@nus.edu.sg}  

\address[CEE]{Department of Civil and Environmental Engineering, National University of Singapore} 
\address[Amherst]{Department of Computer Science, Amherst College}

\address[IORA]{Institute of Operations Research and Analytics, National University of Singapore}


\begin{abstract}                         
Ensuring string stability is critical for the safety and efficiency of large-scale interconnected systems. Although learning-based controllers (e.g., ones based on reinforcement learning) have demonstrated strong performance in complex control scenarios, their black-box nature hinders formal guarantees of string stability. To address this gap, we propose a novel verification and synthesis framework that integrates discrete-time scalable input-to-state stability (sISS) with neural network verification to formally guarantee string stability in interconnected systems. Our contributions are four-fold. First, we establish a formal framework for synthesizing and robustly verifying discrete-time scalable input-to-state stability (sISS) certificates for neural network–based interconnected systems. Specifically, our approach extends the notion of sISS to discrete-time settings, constructs neural sISS certificates, and introduces a verification procedure that ensures string stability while explicitly accounting for discrepancies between the true dynamics and their neural approximations. Second, we establish theoretical foundations and algorithms to scale the training and verification pipeline to large‑scale interconnected systems. Third, we extend the framework to handle systems with external control inputs, thereby allowing the joint synthesis and verification of neural certificates and controllers.
Fourth, we validate our approach in scenarios of mixed-autonomy platoons, drone formations, and microgrids. Numerical simulations show that the proposed framework not only guarantees sISS with minimal degradation in control performance but also efficiently trains and verifies controllers for large‑scale interconnected systems under specific practical conditions. 
\end{abstract}

\end{frontmatter}

\section{Introduction}
\label{sec:intro}
Large-scale interconnected systems~\cite{haber2014subspace,lyu2022small,zhang2025privacy,janssen2024modular} arise in many real-world applications, such as power grids~\cite{silva2021string}, smart transportation~\cite{wang2022optimal,zhou2024parameter}, and industrial process control~\cite{zhu2002model}. These systems have become increasingly complex and tightly coupled due to advances in sensing and communication technologies. A critical challenge in controlling interconnected systems is to ensure \emph{string stability}, which prevents local disturbances from amplifying when propagating through the network~\cite{feng2019string}. In vehicle platoon systems, for example, the loss of string stability can cause minor speed fluctuations to escalate into stop-and-go waves, thereby degrading efficiency and compromising safety.

Control strategies for enhancing string stability in interconnected systems can be divided into model-based and learning-based approaches. Model-based controllers, such as linear feedback control~\cite{wang2021leading}, model predictive control~\cite{gratzer2022string,aboudonia2025adaptive}, and sliding mode control~\cite{guo2016distributed}, enable direct analysis of string stability using time-domain (Lyapunov-based) or frequency-domain (transfer function-based) techniques~\cite{feng2019string}. However, these model-based methods often rely on accurate system models and can become less effective or overly conservative when dealing with complex or uncertain dynamics in large-scale interconnected systems. 
To address these limitations, learning-based controllers, in particular, neural network-based controllers, have gained increasing popularity thanks to their ability to handle complex dynamics and uncertain environments in interconnected systems~\cite{cheng2019end,li2021reinforcement,zhou2024enhancing}. 
However, due to the black-box nature of neural networks, it is challenging to provide formal guarantees for string stability. For existing neural network-based controllers, string stability is often treated as a soft constraint in the training process (e.g., incorporated into the loss or reward function) without offering a rigorous theoretical guarantee.

To the best of our knowledge, few studies have addressed the challenge of verifying and ensuring string stability in neural network-based controllers. The only notable attempt is the Estimation-Approximation-Derivation-Calculation framework proposed by Zhang et al. \cite{zhang2024string} that approximates learning-based car-following controllers as linear models and then uses transfer function analysis to assess string stability. However, this approach suffers from significant approximation errors, making it unsuitable for rigorous string stability verification. 
It is worth noting that several works seek to provide formal guarantees for \emph{local stability} rather than string stability. For instance, Dai et al.~\cite{dai2021lyapunov} propose a Lyapunov-based approach to ensure local stability by designing neural network controllers with a verifiable Lyapunov certificate. Building on this, Yang et al.~\cite{yanglyapunov} introduce a framework for verifying neural control under both state and output feedback, further extending Lyapunov-based guarantees. Moreover, Mandal et al.~\cite{mandal2024safe,mandal2024formally} leverage Lyapunov barrier certificates to formally verify deep reinforcement learning controllers, demonstrating safe and reliable training for aerospace applications. Zhang et al.~\cite{zhang2023compositional} train compositional Lyapunov certificates for networked systems and introduce strategies to simplify the training process.
However, these works on local stability cannot be readily extended to ensure string stability in interconnected systems. 
First, local stability guarantees that small perturbations around an equilibrium for an individual agent decay, whereas string stability ensures that disturbances do not amplify as they propagate along a chain of agents. Thus, local stability alone is insufficient for analyzing the cumulative effects of inter-agent disturbance propagation inherent in string stability.
Second, local stability analyses typically focus on small, isolated systems, whereas string stability requires analysis of large-scale networks, presenting scalability challenges. 
Consequently, formally guaranteeing string stability within interconnected systems with neural network controllers remains an open and challenging problem.

\emph{Statement of Contribution}. To bridge the research gap, we propose a scalable verification and synthesis framework for learning a neural controller with a formal string stability guarantee. Our contributions are four-fold. First, we establish a formal framework for the synthesis and robust verification of discrete-time scalable input-to-state stability (sISS) certificates for neural network-based interconnected systems. Specifically, we extend the notion of sISS~\cite{silva2024scalable} to discrete-time systems, construct neural sISS certificates, and develop a robust verification procedure that guarantees string stability while explicitly accounting for discrepancies between the true dynamics and their neural representations.
Second, we address the scalability challenge by developing theoretical foundations and algorithms that reuse neural sISS certificates across smaller or structurally equivalent systems, thereby significantly improving the efficiency of verification and training for large-scale interconnected systems. Third, we generalize the framework to systems with external control inputs, enabling the simultaneous synthesis and verification of neural certificates and controllers. Fourth, we further validate the approach in multiple interconnected systems, including mixed-autonomy platoons, drone formations, and microgrids. Simulation results demonstrate that our framework ensures discrete-time sISS in large-scale systems while preserving control performance.

\section{Related Work}
This work contributes to two fields of study: (i) string stable control for interconnected systems and (ii) neural certificates and neural network verification.

\textbf{String Stable Control.} 
String stability is an important property in the analysis of interconnected systems, which ensures that external disturbances do not amplify as they propagate through successive agents~\cite{feng2019string,rodonyi2019heterogeneous,silva2024scalable}. This concept has been extensively studied in various applications, such as vehicular platoons~\cite{wang2021leading,silva2021string2,vargas2025stochastic}, drone formations~\cite{riehl2022string}, and microgrids~\cite{silva2021string}. Most existing works focus on analyzing string stability for model-based controllers~\cite{feng2019string}, such as linear control~\cite{wang2021leading}, model predictive control~\cite{gratzer2022string}, and $H_\infty$ control~\cite{kayacan2017multiobjective}. In contrast, few works have investigated string stability for model-free controllers, particularly neural network-based controllers~\cite{zhou2024enhancing}. Among the few existing works, Zhang et al.~\cite{zhang2024string} relies on linearization and may make inaccurate conclusions, and Zhou et al.~\cite{zhou2023data} remains largely empirical and does not offer theoretical guarantees. Moreover, existing works are largely restricted to vehicle platoons with simple topologies, and their results may not generalize to more complex, large-scale interconnected systems. To the best of the authors’ knowledge, there is currently no research that provides a rigorous guarantee of string stability for neural network-based controllers.

\textbf{Neural Certificates and Neural Network Verification}.
A growing body of research has examined the use of neural certificates~\cite{dawson2023safe} to embed control properties (e.g., safety and stability) directly into neural network-based controllers. Two prominent examples are neural barrier certificates~\cite{yang2023model,zhao2020synthesizing,10886052,10591251,zhang2023exact,salamati2024data,abate2024safe,pmlr-v270-hu25a,zhang2025gcbf+} and neural Lyapunov certificates~\cite{dai2021lyapunov,zhang2023compositional,yanglyapunov,berger2023counterexample,debauche2024stability,mandal2024formally,liu2025physics}. To construct such certificates, various training techniques have been proposed. For instance, certified training~\cite{mueller2023certified} and adversarial training~\cite{tramer2020adaptive} aim to produce neural networks that empirically satisfy the desired certificate conditions. However, these methods are often overly conservative or fail to offer formal guarantees. 
In contrast, combining such certificate-based methods with verification tools such as $\alpha,\beta$-CROWN~\cite{wang2021beta}, Marabou~\cite{wu2024marabou}, and NNV~\cite{lopez2023nnv}, allows for counterexample-guided inductive synthesis (CEGIS)~\cite{ding2022novel,berger2023counterexample,mandal2024formally}, thereby systematically synthesizing controllers that satisfy barrier or Lyapunov properties. Nonetheless, to the best of the authors’ knowledge, no existing work has investigated neural certificates specifically designed to ensure string stability in large-scale interconnected systems.

\section{Scalable Synthesis and Verification of Discrete-Time Scalable Input-To-State Stability}
In this section, we present the proposed scalable framework for verification and synthesis of discrete-time sISS using vector Lyapunov function. Section~\ref{sec: Discrete-time sISS vector Lyapunov Formulation} introduces the theoretical foundations of discrete-time vector sISS Lyapunov functions, including the definition and a sufficient condition. Building on this foundation, Section~\ref{subsec: verification} establishes a neural verification framework for sISS. Section~\ref{subsec: synthesis} then describes the synthesis procedure of neural sISS certificates, which leverages counterexamples identified during verification to train the vector sISS Lyapunov functions. Finally, Section~\ref{subsec:Generalizability Analysis} establishes theoretical results and scalable algorithms that exploit structural properties of specific classes of interconnected systems to maintain tractable training and verification.

\subsection{Discrete-Time sISS Vector Lyapunov  Formulation}
\label{sec: Discrete-time sISS vector Lyapunov  Formulation}
\noindent \textbf{Notation.} We use $\mathbb{R}$ to denote the set of real numbers, 
$\mathbb{R}_{\geq 0}$ the set of nonnegative reals, 
and $\mathbb{N}$ the set of natural numbers.  
For $a \in \mathbb{R}$, $|a|$ denotes the absolute value.  
For $x \in \mathbb{R}^n$, its $p$-norm is denoted by $|x|_p := \Big(\sum_{i=1}^n |x_i|^p\Big)^{1/p}, \quad p \in [1,\infty),$
and $|x|_\infty := \max_{i=1}^n |x_i|$. Given a function $y : \mathbb{N} \to \mathbb{R}^n$,  its $\mathcal{L}_p$-norm is denoted by $\|y\|_{\mathcal{L}_p} := \Big( \sum_{k \in \mathbb{N}} |y_k|_p^p \Big)^{1/p}, 
\quad p \in [1,\infty)$, and its supremum norm by $\|y\|_{\mathcal{L}_\infty} := \sup_{k \in \mathbb{N}} |y_k|_\infty$. The class $\mathcal{K}$ consists of all continuous, strictly increasing functions $\alpha : \mathbb{R}_{\geq 0} \to \mathbb{R}_{\geq 0}$ with $\alpha(0) = 0$. The subclass $\mathcal{K}_\infty$ denotes a subset of $\mathcal{K}$ functions that are unbounded. The class $\mathcal{KL}$ consists of  functions $\beta : \mathbb{R}_{\geq 0} \times \mathbb{N} \to \mathbb{R}_{\geq 0}$ 
such that $\beta(\cdot, k) \in \mathcal{K}$ for each fixed $k \in \mathbb{N}$ 
and $\beta(s, k) \to 0$ as $k \to \infty$ for each fixed $s \geq 0$.  

Consider an interconnected system of $N$ agents, indexed by the set $\mathcal{N} = \{1,\dots,N\}$.
The interconnection topology is represented by an adjacency matrix $G \in \{0,1\}^{N\times N}$, where $G_{i,j} = 1$ indicates that the dynamics of agent $i$ depend on the state of agent~$j$, and $G_{i,j} = 0$ otherwise. 
We represent the system as a tuple $\mathcal{I}=\left(\mathcal{N}, \{\mathcal{E}_i\}_{i\in\mathcal{N}}, \{f_i\}_{i\in\mathcal{N}}\right)$, where $\mathcal{E}_i = \{ j \in \mathcal{N} \mid G_{i,j}=1 \}$ denotes the set of neighbors of agent~$i\in\mathcal{N}$, and $f_i$ denotes its dynamics function. 

The state of agent $i$ evolves according to the discrete-time dynamics
\begin{equation}
\begin{aligned}
x_{i,k+1} = f_{i}\bigl(x_{i,k},\{x_{j,k}\}_{j \in \mathcal{E}_i},d_{i,k}\bigr), k \in \mathbb{N}
\label{eq: system}
\end{aligned}
\end{equation}
where $x_{i,k} \in \mathcal{R}_{i}$ is the state of agent~$i$ at time step $k$, with $\mathcal{R}_{i}\subset \mathbb{R}^{n_i}$ representing its the admissible operating region. $d_{i,k}\in \mathcal{W}_i\subset \mathbb{R}^{p_i}$ is the external disturbance affecting agent $i$ at time step $k$. Let $\mathcal{R}_{\mathcal{E}_i}:=\prod_{j\in\mathcal{E}_i}\mathcal{R}_j$ and $\mathcal{Z}_i:=\mathcal{R}_{i}\times\mathcal{R}_{\mathcal{E}_i}\times\mathcal{W}_i$. 
We assume $\mathcal{Z}_i$ is bounded as required in most physical systems. 
Without loss of generality, set the equilibrium $x_i^*=0$. Note that an analytic form of $f_i(\cdot)$ is not required. Instead, we will later approximate $f_i$ with a data-driven surrogate.

With this local model setup in place, we now turn to the system-level objective of establishing string stability guarantees for the interconnected system.
We aim to synthesize and verify neural certificates that ensure string stability. Intuitively, string stability requires that disturbances \emph{do not amplify} as they propagate through the interconnected system. Def.~\ref{def: siss} formalizes the notion of string stability, namely discrete-time Scalable Input-to-State Stability (sISS), which is extended from~\cite{silva2024scalable}.

\begin{definition}\rm{\textbf{(Discrete-Time Scalable Input-to-State Stability)}}\label{def: siss}
The discrete-time interconnected system in Eq.~\eqref{eq: system} is said to be \emph{sISS} if there exist functions \(\beta \in \mathcal{KL}\) and \(\mu \in \mathcal{K}\), independent of the network size \(N\), such that for any $k \in \mathbb{N}, i \in\mathcal{N}$, any initial state \(x_i(0)\), and any bounded disturbance sequence \(d_i\), the following inequality holds: 
\begin{equation}
\max_{i\in\mathcal{N}}|x_{i,k}|_2 \leq \beta\left(\max_{i\in\mathcal{N}}|x_{i,0}|_2, k\right) + \mu\left( \max_{i\in\mathcal{N}} \|d_i\|_{\mathcal{L}_{\infty}} \right) \label{eq: sISS_def}
\end{equation}
\end{definition}

A sufficient condition for the sISS is given in Theorem~\ref{thm: siss-lyap}. 
\begin{theorem}{\rm{\textbf{(Discrete-Time sISS vector Lyapunov function)}}}
\label{thm: siss-lyap}
The discrete-time interconnected system in Eq.~\eqref{eq: system} is sISS if there exist $\varepsilon\in(0,1)$ and disturbance gain \(\psi \geq 0\), such that 
each agent $i\in\mathcal{N}$ admits a Lyapunov function $V_i: \mathbb{R}^{n_i} \to \mathbb{R}_{\geq 0}$ satisfying the following two conditions: 

\noindent 1. (Class-$\mathcal{K}_\infty$ bounds)  There exists $\alpha_1,\alpha_2 \in \mathcal{K}_\infty$ such that  \begin{equation}
\label{eq: local_sISS_bound}
\alpha_1(|x_{i,k}|_2) \leq V_i(x_{i,k}) \leq \alpha_2(|x_{i,k}|_2),~\forall i \in \mathcal{N}
\end{equation}
\noindent 2. (Decremental conditions) For any $x_{i,k}\in \mathcal{R}_{i},~i\in\mathcal{N}$, there exists positive gains \(\Gamma = \{\gamma_{i,j}\}_{i\in\mathcal{N},j\in\mathcal{E}_i}\)  satisfying the small-gain condition
\begin{equation}
\label{eq: sISS_small_gain}
\max_{i\in \mathcal{N}}  \sum_{j\in\mathcal{E}_i\cup \{i\}} \gamma_{i,j} \le (1-\varepsilon)
\end{equation}
such that the decremental inequality holds for any $i\in \mathcal{N}$:
\begin{equation}
\label{eq: local_sISS_decrement}
V_i(x_{i,k+1}) \leq \sum_{j \in \mathcal{E}_i\cup \{i\}} \gamma_{i,j}V_j(x_{j,k}) + \psi|d_{i,k}|_2
\end{equation}
Then, $\bm{V}=\left[V_1, V_2,\cdots , V_N\right]^\top$ is a discrete-time sISS vector Lyapunov function. 
\end{theorem}

The sufficient condition represented by Theorem~\ref{thm: siss-lyap} can be used to verify the discrete-time sISS property of an interconnected system. To this end, one needs to construct a discrete-time sISS vector Lyapunov function such that each local function $V_i$ is bounded by class-$\mathcal{K}_{\infty}$ functions as in Eqs.~\eqref{eq: local_sISS_bound}-\eqref{eq: local_sISS_decrement}. In Eq.~\eqref{eq: local_sISS_decrement}, the local coupling coefficients $\gamma_{i,j}$ (with $i, j \in \mathcal{N}$) captures the influence of subsystem $j$ on subsystem $i$. 
These coupling constraints make manual construction of vector Lyapunov function difficult, so we propose a neural network approach as presented next. 

\subsection{Verification Formulation}
\label{subsec: verification}
In this subsection, we certify sISS for interconnected systems with neural certificates via neural network verification. We first characterize the system dynamics and Lyapunov functions with neural networks and then verify the sufficient conditions for sISS in Theorem~\ref{thm: siss-lyap}. 

While the conditions in Theorem~\ref{thm: siss-lyap} can be verified on an explicit model of the interconnected system, the exact dynamics $f_i$ are often unknown in practice, and only an approximated model $\tilde{f}_i$ learned from sampled data is available. Therefore, for each agent $i\in\mathcal{N}$, we characterize its system dynamics and Lyapunov function in discrete-time settings as neural networks $\phi_{\mathrm{dyn},i}$ and $\phi_{V_i}$, respectively, written as: 
\begin{align}
\tilde{x}_{i,k+1}&=\tilde{f}_i\bigl(x_{i,k},\{x_{j,k}\}_{j\in\mathcal{E}_i},d_{i,k}\bigr)\notag\\
&=\phi_{\mathrm{dyn},i}\bigl(x_{i,k},\{x_{j,k}\}_{j\in\mathcal{E}_i},d_{i,k}\bigr),\label{eq:approximate system}\\
V_i\bigl(x_{i,k}\bigr)&=\phi_{V_i}\bigl(x_{i,k}\bigr)-\phi_{V_i}\bigl(x_{i}^*\bigr),
\end{align}
where $\tilde{x}_{k+1,i}$ is the next state generated by the approximated model $\tilde{f}_i$, and $x^*_i=0\in\mathcal{R}_{i}$ represents the equilibrium state for agent $i$.

A key challenge is that verifying the Lyapunov decremental condition on the approximated model does not automatically guarantee that the true system also satisfies the condition, due to inevitable sampling and approximation errors.
To bridge this gap, under standard Assumptions~\ref{asmp: Lipschitz} and~\ref{asmp: uniform sample}, we bound the deviation between the true and approximated dynamics, and propagate this deviation to the Lyapunov inequality using Lipschitz continuity.

\begin{assumption}[Lipschitz Continuity]
\label{asmp: Lipschitz}
The true dynamics $f_i$ in Eq.~\eqref{eq: system} and approximated dynamics $\tilde{f}_i$ in Eq.~\eqref{eq:approximate system} are Lipschitz continuous with constants $L_{f_i}$ and $L_{\tilde{f}_i}$, respectively. The Lyapunov functions $V_i$ are Lipschitz continuous with constant $L_{V_i}$. 
\end{assumption}

\begin{assumption}[Bounded Approximation Error]
\label{asmp: uniform sample}
Given a bounded region of interest $\mathcal{Z}_i \subset \mathbb{R}^{m}$ for agent $i$ with $m = n_i + \sum_{j\in\mathcal{E}_i} n_j + p_i$, let $z_{i,k} := (x_{i,k},\{x_{j,k}\}_{j\in\mathcal{E}_i},d_{i,k})\in\mathcal{Z}_i$ collect the local state, neighbor states, and disturbance. 
Construct a dataset $\mathcal{D}_i \subset \mathcal{Z}_i$ by discretizing each coordinate of $\mathcal{Z}_i$ on a rectangular grid with per-dimension step sizes $\boldsymbol{\Delta}=(\Delta_1,\dots,\Delta_m)\in\mathbb{R}^m_{>0}$. 
Let $\hat{\epsilon}_i$ denote the empirical maximum approximation error of the surrogate $\tilde f_i$ with respect to the true dynamics $f_i$ on $\mathcal{D}_i$, written as
\begin{align}
\hat{\epsilon}_i \;:=\; \max_{z_{i,k}\in \mathcal{D}_i} \, | f_i(z_{i,k}) - \tilde{f}_i(z_{i,k}) |_2.
\end{align}
We assume $\hat{\epsilon}_i<+\infty$, meaning that the approximation error is bounded on the discretized grid $\mathcal{D}_i$. 
\end{assumption}
Under Assumptions~\ref{asmp: Lipschitz} and~\ref{asmp: uniform sample}, the mismatch between the true and approximated models can be bounded explicitly. The following theorem shows that if the Lyapunov decremental condition in Eq.~\eqref{eq: new_siss_decremental} holds for the approximated model with an additional margin accounting for this error bound, the original system is guaranteed to be sISS.

\begin{theorem}[Robust sISS Verification]
\label{thm: verify_neural_network}
Suppose Assumptions~\ref{asmp: Lipschitz} and~\ref{asmp: uniform sample} hold for an interconnected system with true dynamics $x_{i,k+1} = f_i(x_{i,k}, \{x_{j,k}\}_{j \in \mathcal{E}_i}, d_{i,k})$ and approximated dynamics $\tilde{x}_{i,k+1} = \tilde{f}_i(x_{i,k}, \{x_{j,k}\}_{j \in \mathcal{E}_i}, d_{i,k})$. Assume each agent $i\in\mathcal{N}$ admits a Lyapunov function $V_i$ satisfying class-$\mathcal{K}_\infty$ bounds in Eq.~\eqref{eq: local_sISS_bound}. 
Then, the true system is discrete-time sISS if these Lyapunov functions satisfy the following decremental condition for the approximated system, i.e., for any $\tilde{x}_{i,k}\in \mathcal{R}_i,~i\in\mathcal{N}$, there exist positive gains \(\Gamma = \{\gamma_{i,j}\}_{i\in\mathcal{N},j\in\mathcal{E}_i}\) satisfying Eq.~\eqref{eq: sISS_small_gain} such that 
\begin{equation}
V_i(\tilde{x}_{i,k+1}) \leq \sum_{j \in \mathcal{E}_i \cup \{i\}} \gamma_{i,j}V_j(x_{j,k}) +\psi|d_{i,k}|_2 - \delta_i,\label{eq: new_siss_decremental}
\end{equation}
with $\delta_i \geq L_{V_i} \epsilon_i > 0$, $\epsilon_i = \hat{\epsilon}_i + \frac{1}{2}(L_{f_i} + L_{\tilde{f}_i}) |\boldsymbol{\Delta}|_2$.
\end{theorem}

\begin{remark}
On the operating set $\mathcal Z_i$, the Lipschitz constants used in Theorem~\ref{thm: verify_neural_network} are instantiated as follows:
(i)~$L_{f_i}$ of the true dynamics is assumed known a priori from physics/regularity of $f_i$ on $\mathcal Z_i$ and treated as a fixed constant; 
(ii)~$L_{\tilde f_i}$ and $L_{V_i}$ are estimated via a neural Lipschitz bounding method such as~\cite{xu2024eclipse}, which provides sound upper bounds for feed-forward networks on a given domain.
\end{remark}

Theorem~\ref{thm: verify_neural_network} provides sufficient conditions for guaranteeing sISS of the original system in terms of the local Lyapunov function inequalities. 
These conditions are stated for each agent $i\in\mathcal{N}$ and depend only on the states and inputs within its local neighborhood $\mathcal{E}_i$. 
This locality naturally enables a distributed verification strategy, where each agent can independently check the validity of Eq.~\eqref{eq: local_sISS_bound} and Eq.~\eqref{eq: new_siss_decremental} without requiring global state information. The global sISS certificate is then constructed as:
\begin{equation}
    \bigwedge_{i \in \mathcal{N}}\Bigl[\text{Eq.}~(\ref{eq: local_sISS_bound}) \land \text{Eq.}~(\ref{eq: new_siss_decremental})\Bigr].
    \label{eq: verification for all agents}
\end{equation}
where $\bigwedge$ encodes the joint satisfaction of all agent conditions, i.e., the logical \textsc{and} of all agent requirements.
Eq.~\eqref{eq: verification for all agents} decomposes a large verification query on a complex multi-agent system into a number of smaller verification queries. Neural network verification tools like $\alpha$-$\beta$-crown~\cite{wang2021beta} and Marabou~\cite{wu2024marabou} can be utilized to check Eq.~\eqref{eq: verification for all agents}.

\subsection{Synthesis of Neural sISS certificates}
\label{subsec: synthesis}
Guided by the sufficient conditions in Theorem~\ref{thm: verify_neural_network}, we jointly search for the coupling matrix and vector Lyapunov candidates to facilitate training. 
To this end, we parameterize the coupling matrix $\Gamma$ by combining a learnable matrix $\Gamma_{\text{pure}}$ and adjacency matrix $G$ as $\Gamma = \operatorname{ReLU}(\Gamma_{\text{pure}}) \circ G$, where the elementwise product $\circ$ ensures $\gamma_{i,j} = 0$ when $G_{i,j} = 0$, and the ReLU activation guarantees $\gamma_{i,j} \geq 0$. To satisfy the small-gain condition by construction, we compute $s_i = \sum_{j\in \mathcal E_i \cup {i}} \gamma_{i,j}$ and rescale each row by $\gamma_{i,j} \;\leftarrow\; \frac{1-\varepsilon}{s_i}\,\gamma_{i,j}, \quad j \in \mathcal E_i \cup \{i\}$. The matrix $\Gamma_{\text{pure}}$ is then trained together with Lyapunov functions to identify the matrix $\Gamma$. 

We synthesize the neural sISS certificates using the following loss function:
\begin{align}
    & \mathbb{L}(\Gamma,\boldsymbol{V}) = w_{\text{p}}\mathbb{L}_{\text{p}} + w_{\text{d}}\mathbb{L}_{\text{d}}  \label{eq: total_loss} \\
    & \mathbb{L}_{\text{p}} = \frac{1}{|\mathcal{N}|}\sum_{i\in \mathcal{N}}\text{ReLU}\left(\alpha_1(|x_{i,k}|_2)-V_i(x_{i,k})+\epsilon_{\text{p}}\right)\notag\\
    &+\text{ReLU}\left(-\alpha_2(|x_{i,k}|_2)+V_i(x_{i,k})+\epsilon_{\text{p}}\right),\label{eq: loss_obj3}\\
    & \mathbb{L}_{\text{d}} = \frac{1}{\lvert\mathcal{N}\lvert}\sum_{i\in \mathcal{N}}\text{ReLU}(V_i(x_{i,k+1})-\gamma_{i,i}V_i(x_{i,k})\notag\\
&-\sum_{j\in\mathcal{E}_i} \gamma_{i,j}V_j(x_{j,k})
-\psi\lvert d_{i,k}\rvert_2+\delta_i+\epsilon_{\text{d}}),\label{eq: loss_obj4}
\end{align}
where the total loss $\mathbb{L}(\Gamma,\boldsymbol{V})$ includes two components: (i) $\mathbb{L}_{\text{p}}$ enforces the class-$\mathcal{K}_{\infty}$ bounds in Eq.~\eqref{eq: loss_obj3} and (ii) $\mathbb{L}_{\text{d}}$ enforces the decrement inequality in Eq.~\eqref{eq: loss_obj4}. The coefficients $w_{\text{p}}, w_{\text{d}}$
are weighting factors that balance the different objectives in the loss function. The values $\epsilon_{\text{p}}$ and $\epsilon_{\text{d}}$ act as margins to ensure that we can adopt a more conservative policy to satisfy the sISS conditions.

With the training and verification formulation, we use a counterexample-guided inductive synthesis (CEGIS) loop to obtain a neural sISS certificate as in Algorithm~\ref{alg:joint-training-cegis}. At each CEGIS iteration, we jointly train $\{V_i\}_{i\in\mathcal{N}}$ and $\Gamma$ to minimize the loss in Eq.~(\ref{eq: total_loss}) and then use a formal verifier (e.g., \cite{nnvTwo,wu2024marabou,wang2021beta,xu2020automatic,verinet}) to verify the certificate. If the verifier returns a counterexample violating constraints Eq.~(\ref{eq: local_sISS_bound}) or Eq.~(\ref{eq: new_siss_decremental}), we sample points in a small neighborhood of that counterexample and add them to the training set. The reason for sampling multiple points near the counterexamples is to learn smooth behavior in the neighborhood instead of overfitting to a specific point. This process is repeated iteratively until no counterexamples are found, yielding a fully verified neural sISS vector Lyapunov certificates.

\begin{algorithm}[htp]
\caption{Counterexample-Guided Inductive Synthesis Loop for sISS Certificate}
\label{alg:joint-training-cegis}
\begin{algorithmic}[1]
\Require Initial dataset $D$ containing state-action pairs, initial parameters of Lyapunov functions $\{V_i\}_{i\in\mathcal{N}}$, an initial coupling matrix $\Gamma$, and a neural network verifier $\text{Ver}_{\text{NN}}$.
\Ensure Verified vector Lyapunov function $\{V_i\}_{i\in\mathcal{N}}$, and coupling matrix $\Gamma$.
\Repeat
    \State Train $\Gamma$ and $\{V_i\}_{i\in\mathcal{N}}$ by minimizing the loss function~Eq.~(\ref{eq: total_loss}).
    \State Verify the sISS conditions~Eq.~(\ref{eq: local_sISS_bound}) and~Eq.~(\ref{eq: new_siss_decremental}) using the neural network verifier $\text{Ver}_{\text{NN}}$.
    \If{counterexamples violating the constraints are found}
        \State Augment dataset $D$ with counterexamples.
    \EndIf
\Until{no counterexamples are found after verification.}
\end{algorithmic}
\end{algorithm}

\subsection{Scalability Analysis}
\label{subsec:Generalizability Analysis}
The proposed framework for synthesising neural sISS certificates can be computationally challenging for large-scale interconnected systems, due to the NP-hardness of finding a verified neural sISS certificate and the slow convergence rate of the counterexample-guided training process. To address this, we develop a scalable synthesis framework that accelerates certificate synthesis by reusing the neural sISS certificates for smaller or structurally equivalent systems. We next present theoretical results supporting the synthesis framework. 

\textbf{(1) Structural properties.} We show in Theorem~\ref{thm:Certifiable Equivalence} that sISS certificates for a class of large interconnected systems can be reconstructed from those of smaller systems without requiring re-verification, under certain structural conditions formalized in Def.~\ref{dfn:stru_eq}, i.e., all ``substructures'' in $\tilde{\mathcal{I}}$ can be found in $\mathcal{I}$.

\begin{definition}[Substructure Isomorphism] \label{dfn:stru_eq}
An interconnected system $\tilde{\mathcal{I}} =(\widetilde{\mathcal{N}}, \{\widetilde{\mathcal{E}}_j\}_{j\in\widetilde{\mathcal{N}}}, \{\widetilde{f}_j\}_{j\in\widetilde{\mathcal{N}}})$ is \textbf{substructure-isomorphic} to $\mathcal{I} = (\mathcal{N}, \{\mathcal{E}_i\}_{i\in\mathcal{N}}, \{f_i\}_{i\in\mathcal{N}})$, if there exists an injective map $\tau : \widetilde{\mathcal{N}} \rightarrow \mathcal{N}$ satisfying $\tau(\widetilde{\mathcal{E}}_j) = \mathcal{E}_{\tau(j)}$ and $\widetilde{f}_j = f_{\tau(j)}$ for each $j \in \widetilde{\mathcal{N}}$.     
\end{definition}

\begin{theorem}{\rm{\textbf{(sISS Preservation under Substructure Isomorphism)}}}
\label{thm:Certifiable Equivalence}
An interconnected system $\mathcal{I} = (\mathcal{N}, \{\mathcal{E}_i\}_{i\in\mathcal{N}}, \{f_i\}_{i\in\mathcal{N}})$ admits sISS certificates $\{V_i\}_{i\in\mathcal{N}}$ satisfying the conditions of Eqs.~\eqref{eq: local_sISS_bound}--\eqref{eq: local_sISS_decrement} in Theorem~\ref{thm: siss-lyap}. Then, any interconnected system $\tilde{\mathcal{I}} =(\widetilde{\mathcal{N}}, \{\widetilde{\mathcal{E}}_j\}_{j\in\widetilde{\mathcal{N}}}, \linebreak\{\widetilde{f}_j\}_{j\in\widetilde{\mathcal{N}}})$ substructure-isomorphic to $\mathcal{I}$ admits 
sISS certificates $\{\widetilde{V}_i\}_{i\in\widetilde{\mathcal{N}}}$, where  $\widetilde{V}_j = V_{\tau(j)},~\forall j \in \widetilde{\mathcal{N}}$. 
\end{theorem}

Similar properties can be generalized to each node in Def.~\ref{dfn:eq_nodes} to describe the structural equivalence between nodes (see Appendix~\ref{appendix: example 2} for an illustrative example). We show in Theorem~\ref{thm:eq_nodes} that structurally equivalent nodes share the same certificate, which can significantly reduce verification complexity. 
\begin{definition}[Node Structural Equivalence] \label{dfn:eq_nodes}
    Let $\mathcal{I} = (\mathcal{N}, \{\mathcal{E}_i\}_{i\in\mathcal{N}}, \{f_i\}_{i\in\mathcal{N}})$ be an interconnected system. If there exists a permutation $\tau$ of $\mathcal{N}$ such that $\tau(\mathcal{E}_j) = \mathcal{E}_{\tau(j)}$ and $f_j = f_{\tau(j)}$, $\forall j \in \mathcal{N}$, node $j$ is said to be \textbf{structurally equivalent} to node $\tau(j)$ for each $j \in \mathcal{N}$. 
\end{definition}

\begin{theorem}{\rm{\textbf{(Identical Certificates for Structural Equivalent Nodes)}}}\label{thm:eq_nodes}
    Suppose an interconnected system $\mathcal{I} = (\mathcal{N}, \{\mathcal{E}_i\}_{i\in\mathcal{N}}, \{f_i\}_{i\in\mathcal{N}})$ satisfies the conditions of Eqs.~\eqref{eq: local_sISS_bound}--\eqref{eq: local_sISS_decrement} in Theorem~\ref{thm: siss-lyap}. Then, there exists sISS certificates $\{V_i\}_{i\in\mathcal{N}}$ such that $V_i = V_{j}$ for any structurally equivalent nodes $i,j \in \mathcal{N}$. 
\end{theorem}

Theorem~\ref{thm:eq_nodes} suggests that the node set $\mathcal{N}$ can be partitioned into structurally equivalent classes according to Def.~\ref{dfn:eq_nodes}. All nodes in the same equivalent classes have the same certificates. Therefore, we can simplify the verification process by verifying one representative node in each equivalent class. 

\textbf{(2) Generalizability.} Theorems~\ref{thm:Certifiable Equivalence}-\ref{thm:eq_nodes} require the dynamics of corresponding agents to be exactly the same, which may not hold in practice. To accommodate the discrepancies in system dynamics, Theorem~\ref{thm:siss_affine_last_layer} establishes conditions under which sISS certificates generalize across interconnected systems with dynamics affine in a parameter vector under the assumption that the Lyapunov function is convex (e.g., by using the input convex neural network~\cite{amos2017input}). Specifically, sISS certificates verified at the vertices of a convex parameter space remain valid throughout the entire space. This supports practical generalization, as neural network-based dynamics models can adapt to new systems by fine-tuning only the last-layer parameters. A practical algorithm for applying Theorem~\ref{thm:siss_affine_last_layer} in the real world is given in Appendix~\ref{appendix: Algorithm for Theorem siss_affine}.

\begin{theorem}{\rm{\textbf{(sISS with Parameter-Affine Dynamics)}}}
\label{thm:siss_affine_last_layer}
Let $\mathcal{B}\subset \mathbb{R}^{p}$ be the convex hull of a finite set of parameter vectors
$\{\chi^{\nu}\}_{\nu\in\Omega}$. Consider an interconnected system in Eq.~\eqref{eq: system}, where each agent $i\in\mathcal{N}$ evolves according to dynamics parametrized by $\chi\in\mathcal B$:  
\begin{align}    
x_{i,k+1}
& =
f_i\bigl(x_{i,k},\{x_{j,k}\}_{j\in\mathcal E_i},d_{i,k};\chi\bigr) \notag \\ 
& = \sum_{l=1}^p\chi_l\Phi_{i,l}\bigl(x_{i,k}, \{x_{j,k}\}_{j\in\mathcal E_i}, d_{i,k}\bigr),
\label{eq: system_affine}
\end{align}
where the parametrization is affine in $\chi$, with $\Phi_{i,l}$ fixed and $\chi$-independent for $l=1,\cdots,p$.

Assume each $V_i:\mathbb{R}^{n_i}\to\mathbb{R}_{\ge0}$ is convex.  
Suppose there exist a positive scalar $\psi$  and positive gains \(\Gamma = \{\gamma_{i,j}\}_{i\in\mathcal{N},j\in\mathcal{E}_i}\) satisfying the small gain condition Eq.~\eqref{eq: sISS_small_gain} such that for every vertex $\chi^{\nu}$, the decremental condition holds for all admissible states and disturbances:  
\begin{equation}\label{eq:local_siss_vertex}
\begin{aligned}
&V_i\bigl(x_{i,k+1};\chi^{\nu}\bigr)
\le\sum_{j\in\mathcal E_i\cup \{i\}}\gamma_{i,j}V_j(x_{j,k})
+\psi|d_{i,k}|_2,
\ \forall i\in\mathcal N.
\end{aligned}
\end{equation}
Then for any $\chi\in\mathcal{B}$, the same vector Lyapunov function $\{V_i\}_{i\in\mathcal N}$ verifies
Eq.~\eqref{eq:local_siss_vertex}, and the interconnected system Eq.~\eqref{eq: system_affine} is sISS.
\end{theorem}

\textbf{(3) Modular verification.} We consider another class of large-scale interconnected systems that admits modular verification: if the original network is already verified, and an added sub-network is verified in isolation with properly accounted couplings, then the combined network remains verified. 

To formalize this principle, we first define a subset relation between two interconnected systems.
For interconnected systems, the subset relation $\mathcal{I}' = (\mathcal{N}', \{\mathcal{E}'_i\}_{i\in\mathcal{N}'}, \{f_i'\}_{i\in\mathcal{N}'}) \subset\mathcal{I} =(\mathcal{N}, \{\mathcal{E}_i\}_{i\in\mathcal{N}}, \{f_i\}_{i\in\mathcal{N}})$ implies (i) $\mathcal{N}'\subset \mathcal{N}$, (ii) $\mathcal{E}'_i = \{(i,j) \in \mathcal{E}_i  \mid j\in\mathcal{N}'\}$, and (iii) $f_i'\bigl(x_{i},\{x_j\}_{j \in \mathcal{E}_i'},d_i\bigr)=f_{i}\bigl(\tilde{x}_{i},\{\tilde{x}_j\}_{j \in \mathcal{E}_i},d_i\bigr)$, $\forall i \in \mathcal{N'}$, where $\tilde{x}_j=x_j$ if $j\in\mathcal{N}'$ and $\tilde{x}_j=0$ otherwise.  

Building on this relation, we next define the \emph{decomposability} of interconnected systems in Def.~\ref{dfn:additiveness} that supports modular verification, and then present sufficient conditions in Theorem~\ref{thm:augmentedSystem}. An illustrative example is given in Appendix~\ref{appendix: example 1}, which can be seen as a vehicle platoon where each following vehicle relies only on the state of its immediate predecessor.

\begin{definition}[sISS Decomposability] \label{dfn:additiveness}
An interconnected system $\mathcal{I} = (\mathcal{N}, \{\mathcal{E}_i\}_{i \in \mathcal{N}}, \{f_i\}_{i\in\mathcal{N}})$ is said to be \textbf{sISS decomposable} 
if it contains a subsystem 
$\mathcal{I}' = (\mathcal{N}', \{\mathcal{E}'_i\}_{i \in \mathcal{N}'}, \{f_i'\}_{i\in\mathcal{N}'})\subset\mathcal{I}$ that is already sISS-verified, and verifying sISS for the remaining agents in $\mathcal{N}\backslash\mathcal{N}'$ suffices to guarantee sISS for the full system  $\mathcal{I}$.

\end{definition}

\begin{theorem}{\rm{\textbf{(Sufficient Condition for sISS Decomposability)}}}
Consider an interconnected system $\mathcal{I} = (\mathcal{N}, \{\mathcal{E}_i\}_{i \in \mathcal{N}}, \{f_i\}_{i\in\mathcal{N}})$ containing a subsystem $\mathcal{I}' = (\mathcal{N}', \{\mathcal{E}'_i\}_{i \in \mathcal{N}'}, \{f_i'\}_{i\in\mathcal{N}'})\subset\mathcal{I}$ that is already sISS-verified with certificates $\{V_i'\}_{i \in \mathcal{N}'}$ satisfying the conditions of Eqs.~\eqref{eq: local_sISS_bound}--\eqref{eq: local_sISS_decrement} in Theorem~\ref{thm: siss-lyap}. The system $\mathcal{I}$ is sISS decomposable if the subsystem dynamics are independent of the remaining states of other agents, i.e., $f_i(x_{i,k},\{x_{j,k}\}_{j\in \mathcal{E}_i},d_{i,k})=f_i'(x_{i,k},\{x_{j,k}\}_{j\in \mathcal{E}_i'},d_{i,k}),~\forall i \in \mathcal{N}'$. 
\label{thm:augmentedSystem}
\end{theorem}

Overall, Theorems \ref{thm:Certifiable Equivalence} -- \ref{thm:augmentedSystem} enable us to reuse certificates derived from smaller and structurally equivalent systems. Building on the proposed theorems, we devise an algorithm that reduces a large interconnected system to a minimum verification network (see Alg.~\ref{alg:mvn-reduction} in Appendix~\ref{appendix:Algorithm for Constructing Minimum Verification Network}). In the next section, we extend our framework to systems with external control inputs, enabling the simultaneous synthesis and verification of neural certificates and controllers.

\section{Scalable Synthesis and Verification of Discrete-Time sISS under Control}
In this section, we extend the framework to systems with external control inputs. In this setting, stability is no longer guaranteed \emph{a priori} but must be enforced through the design of suitable controllers. To this end, we introduce the notion of discrete-time sISS vector control Lyapunov functions (sISS-VCLFs) as in Section~\ref{subsec: siss_lyap}, which characterize 
the existence of feedback laws ensuring scalable stability. Based on this formulation, we develop synthesis and verification procedures that allow the simultaneous construction of neural certificates and controllers in Section~\ref{subsec: Synthesis and Verification Framework with control}, and scalability analysis in Section~\ref{subsec: Scalability Analysis Control}.

\subsection{Discrete-Time sISS Vector Control Lyapunov Functions}
\label{subsec: siss_lyap}
Consider the interconnected system with external control input:
\begin{equation}
\begin{aligned}
x_{i,k+1} = f_{i}\bigl(x_{i,k},\{x_{j,k}\}_{j \in \mathcal{E}_i},u_{i,k},d_{i,k}\bigr), k \in \mathbb{N},
\label{eq: system_control}
\end{aligned}
\end{equation}
where the control input for agent~$i$ is determined by a neural network-based controller trained from RL (or imitating a traditional controller), written as: 
\begin{equation}
u_{i,k} = \pi_{i}\bigl(x_{i,k},\{x_{j,k}\}_{j \in \mathcal{E}_i}\bigr).
\label{eq: controller}
\end{equation}
Our goal is extended to verify and synthesize neural network-based controllers and certificates for such an interconnected system so that string stability is guaranteed. 

For systems with control inputs, Theorem~\ref{thm: siss-lyap} directly leads to the following corollary on vector control Lyapunov functions. It formalizes that, whenever such functions and corresponding feedback laws exist, the closed-loop interconnected system is guaranteed to satisfy the discrete-time sISS property.
\begin{corollary}[Close-loop extension of Theorem~\ref{thm: siss-lyap}]
\label{col: disc_siss_vclf}
Consider the discrete-time interconnected system Eq.~\eqref{eq: system_control}. 
If there exist locally Lipschitz continuous feedback laws $\{\pi_i\}_{i\in\mathcal{N}}$ 
and a family of vector Lyapunov function $\{V_i\}_{i\in\mathcal{N}}$ that satisfies the conditions in Theorem~\ref{thm: siss-lyap}, 
then the closed-loop system Eq.~\eqref{eq: system_control} is discrete-time 
sISS according to Definition~\ref{def: siss}.
\end{corollary}

\subsection{Synthesis and Verification Framework}
\label{subsec: Synthesis and Verification Framework with control}
Similarly, we model the system dynamics, control policy, and Lyapunov functions as neural networks as follows:
\begin{align}
\tilde{x}_{i,k+1}&=\tilde{f}_i\bigl(x_{i,k},\{x_{j,k}\}_{j\in\mathcal{E}_i},u_{i,k},d_{i,k}\bigr)\notag\\
&=\phi_{\mathrm{dyn},i}\bigl(x_{i,k},\{x_{j,k}\}_{j\in\mathcal{E}_i},u_{i,k},d_{i,k}\bigr),\label{eq:approximate system control}\\
u_{i,k}&=\pi_i\bigl(x_{i,k},\{x_{j,k}\}_{j \in \mathcal{E}_i}\bigl)\nonumber\\
&=\operatorname{clamp}\Bigl(\phi_{\pi_i}\bigl(x_{i,k},\{x_{j,k}\}_{j \in \mathcal{E}_i}\bigl),\,u_{\min},\,u_{\max}\Bigr),\\
V_i\bigl(x_{i,k}\bigr)&=\phi_{V_i}\bigl(x_{i,k}\bigr)-\phi_{V_i}\bigl(x_{i}^*\bigr),
\end{align}
where $u_{\text{min}}$ and $u_{\text{max}}$ are the controller bounds, and the $\operatorname{clamp}$ function ensures that the output of $\phi_{\pi_i}$ is restricted within the range $[u_{\text{min}},u_{\text{max}}]$. $x^*_i=0\in\mathcal{R}_{i}$ represents the equilibrium state for agent $i$.  Moreover, we represent the pre-trained neural network-based policy, which requires further verification and fine-tuning, as $\pi_{i,\text{ori}}$. 

Then we synthesize the neural sISS certificates and controllers using the following loss function:
\begin{align}
    & \mathbb{L}_{\text{o}} = \frac{1}{|\mathcal{N}|}\sum_{i\in \mathcal{N}}| \pi_i(x_{i,k}) - \pi_{i,\text{ori}}(x_{i,k})|_2,\label{eq: loss_obj1}\\
    & \mathbb{L}(\Gamma,\boldsymbol{\pi},\boldsymbol{V}) = w_{\text{o}}\mathbb{L}_{\text{o}} +  w_{\text{p}}\mathbb{L}_{\text{p}} + w_{\text{d}}\mathbb{L}_{\text{d}} \label{eq: total_loss_control}
\end{align}
where $\mathbb{L}_{\text{o}}$ represents a controller imitation loss that penalizes the deviation of the synthesized controller $\pi_i$ from the original controller $\pi_{i,\text{ori}}$, it preserves the performance of the original policy. The coefficient 
$w_{\text{o}}$ is the weighting factor for neural network-based controllers. $\mathbb{L}_p,\mathbb{L}_d$ directly follows from Eqs.~\eqref{eq: loss_obj3}-\eqref{eq: loss_obj4}.

Since we aim to verify the controller–plant closed loop, a robust guarantee in the sense of Theorem~\ref{thm: verify_neural_network} requires an extra assumption on the controller and a corresponding extension of Theorem~\ref{thm: verify_neural_network}. The additional assumption and the corollary for robust verification are stated next.

\begin{assumption}{\rm{\textbf{(Lipschitz Continuity for Controllers)}}}
\label{asmp: Lipschitz control}
Each controller $\pi_i$ is Lipschitz continuous with constant $L_{\pi_i}$, $\forall i\in\mathcal{N}$.
\end{assumption}

\begin{corollary}[Close-loop extension of Theorem~\ref{thm: verify_neural_network}]
\label{col: verify_neural_network}
Suppose Assumptions \ref{asmp: Lipschitz}-\ref{asmp: Lipschitz control} hold for an interconnected system with true dynamics $x_{i,k} = f_i(\cdot)$ and approximated dynamics $\tilde{x}_{i,k} = \tilde{f}_i(\cdot)$ with controllers in Eq.~\ref{eq: controller}. Assume each agent $i\in \mathcal{N}$ admits a Lyapunov function $V_i$ satisfying class-$\mathcal{K}_\infty$ bounds in Eq.~\eqref{eq: local_sISS_bound}. Then, the true system is discrete-time sISS if these Lyapunov functions satisfy the following decremental condition for the approximated system, i.e., for any $\tilde{x}_{i,k}\in\mathcal{R}_i,i\in\mathcal{N}$, there exist gains $\Gamma=\{\gamma_{i,j}\}_{i\in\mathcal{N},j\in\mathcal{E}_i}$, satisfying Eq.~\eqref{eq: sISS_small_gain} such that
\begin{equation}
\begin{aligned}
&V_i(\tilde{x}_{i,k+1}) \leq  \sum_{j \in \mathcal{E}_i\cup\{i\}} \gamma_{i,j}V_j({x}_{j,k}) +\psi |d_{i,k}|_2 - \delta_i,\label{eq: new_siss_decremental_control}
\end{aligned}
\end{equation}
with $\delta_i \geq L_{V_i} \epsilon_i > 0$, $\epsilon_i = \hat{\epsilon}_i + \frac{1}{2}(L_{f_i} + L_{\tilde{f}_i})(1 + L_{\pi_i}) |\boldsymbol{\Delta}|_2$.
\end{corollary}

By Corollary 2, the sISS property holds provided the following conjunction of local conditions is satisfied.
\begin{equation}
    \bigwedge_{i \in \mathcal{N}}\Bigl[\text{Eq.}~(\ref{eq: local_sISS_bound}) \land \text{Eq.}~(\ref{eq: new_siss_decremental_control})\Bigr].
    \label{eq: verification for all agents control}
\end{equation}
Using a neural network verifier (e.g., Marabou~\cite{wu2024marabou}), we can similarly construct a CEGIS loop to obtain controllers and Lyapunov functions that satisfy Eq.~\eqref{eq: verification for all agents control}. 

\subsection{Scalability Analysis}
\label{subsec: Scalability Analysis Control}
In this subsection, we extend Theorems~\ref{thm:Certifiable Equivalence}-\ref{thm:augmentedSystem} to systems with control inputs. In particular, Theorem~\ref{thm:Certifiable Equivalence} applies directly to systems with exogenous inputs. For Theorems \ref{thm:eq_nodes}–\ref{thm:augmentedSystem}, we obtain the following corollaries:
\begin{corollary}[Close-loop extension of Theorem~\ref{thm:eq_nodes}]
\label{col: eq_nodes}
    Suppose an interconnected system $\mathcal{I} = (\mathcal{N}, \{\mathcal{E}_i\}_{i\in\mathcal{N}},\linebreak \{f_i\}_{i\in\mathcal{N}})$ satisfies the conditions of Eqs.~\eqref{eq: local_sISS_bound}--\eqref{eq: local_sISS_decrement} in Theorem~\ref{thm: siss-lyap}. Moreover, assume that all agents use the same feedback law, i.e., $\pi_i=\pi$ for all $i\in\mathcal N$. Then, there exists sISS certificates $\{V_i\}_{i\in\mathcal{N}}$ such that $V_i = V_{j}$ for any structurally equivalent nodes $i,j \in \mathcal{N}$. 
\end{corollary}
Since all agents share the same feedback law, the closed-loop system is permutation-invariant with respect to structurally equivalent nodes. Hence Corollary~\ref{col: eq_nodes} follows directly from Theorem~\ref{thm:eq_nodes}.

\begin{corollary}[Close-loop extension of Theorem~\ref{thm:siss_affine_last_layer}]
\label{col: siss_affine_last_layer}
Let $\mathcal{B}\subseteq\mathbb{R}^{p}$ be the convex hull of a finite set of parameter vectors
$\{\chi^{\nu}\}_{\nu\in\Gamma}$, $z_{i,k}=\left(x_{i,k},\{x_{j,k}\}_{j\in\mathcal E_i},d_{i,k}\right)$.
For every $\chi\in\mathcal B$, the $i$-th agent evolves according to control–affine dynamics: 
\begin{align}\label{eq:ca_dynamics}
x_{i,k+1}
&=h_i\bigl(z_{i,k};\chi\bigr)+ g_i\bigl(z_{i,k};\chi\bigr)u_{i,k},\notag\\
&=\sum_{l=1}^{p}\chi_l\left( \Phi_{i,l}^h\bigl(z_{i,k}\bigr)+\Phi_{i,l}^g\bigl(z_{i,k}\bigr)u_{i,k} \right)
\end{align}
where the parametrization is affine in $\chi$, with $\Phi^h_{i,l},\Phi^g_{i,l}$ fixed and $\chi$-independent for $l=1,\cdots,p$.

Assume each $V_i:\mathbb{R}^{n_i}\to\mathbb{R}_{\ge0}$ is convex. Suppose there exist positive gains
$\{\gamma_{i,j}\}_{j\in\mathcal E_i\cup\{i\}},\psi$ satisfying the small-gain condition in Eq.~\ref{eq: sISS_small_gain}, such that for every vertex $\chi^{\nu}$, the decremental condition holds for all admissible states and disturbances,
\begin{equation}\label{eq:local_siss_vertex_control}
\begin{aligned}
V_i\bigl(x_{i,k+1};\chi^{\nu}\bigr)
\le \sum_{j\in\mathcal E_i\cup\{i\}}\gamma_{i,j}V_j(x_{j,k})+\psi |d_{i,k}|_2,
\quad\forall i\in\mathcal N.
\end{aligned}
\end{equation}
Then for any $\chi\in\mathcal{B}$, the same $\{V_i\}_{i\in\mathcal N}$ verify
Eq.~\eqref{eq:local_siss_vertex_control}, and the interconnected system Eq.~\eqref{eq:ca_dynamics} is sISS.
\end{corollary}
Since the affine dependence of $h_i,g_i$ on $\chi$, and the $\chi$-independence of $u_{i,k}$, this corollary directly follows from Theorem~\ref{thm:siss_affine_last_layer}.

\begin{corollary}[Close-loop extension of Theorem~\ref{thm:augmentedSystem}]
Consider an interconnected system $\mathcal{I} = (\mathcal{N}, \{\mathcal{E}_i\}_{i \in \mathcal{N}},\linebreak \{f_i\}_{i\in\mathcal{N}})$ containing a subsystem $\mathcal{I}' = (\mathcal{N}', \{\mathcal{E}'_i\}_{i \in \mathcal{N}'},\linebreak \{f_i'\}_{i\in\mathcal{N}'})\subset\mathcal{I}$ that is already sISS-verified under decentralized controllers $\{\pi_i'\}_{i \in \mathcal{N}'}$ with certificates $\{V_i'\}_{i \in \mathcal{N}'}$ satisfying the conditions of Eqs.~\eqref{eq: local_sISS_bound}--\eqref{eq: local_sISS_decrement} in Theorem~\ref{thm: siss-lyap}. The system $\mathcal{I}$ is decomposable if the subsystem dynamics are independent of the remaining states of other agents, i.e., $f_i(x_{i,k},\{x_{j,k}\}_{j\in \mathcal{E}_i},u_{i,k},d_{i,k})=f_i'(x_{i,k},\{x_{j,k}\}_{j\in \mathcal{E}_i'},u_{i,k},\linebreak d_{i,k}),~\forall i \in \mathcal{N}'$. 
\label{col:augmentedSystem}
\end{corollary}
Since for every $i\in\mathcal N'$, the controller $\pi_i'$ depends only on $(x_i,\{x_j\}_{j\in\mathcal E_i'})$ with $\mathcal E_i'\subset\mathcal N'$ (i.e., it is independent of the states $\{x_j\}_{j\in \mathcal N\setminus\mathcal N'}$), the subsystem certificates remain valid in the augmented system. Hence, Corollary~\ref{col:augmentedSystem} follows directly from Theorem~\ref{thm:augmentedSystem}.

\section{Numerical Simulation}
\label{sec:Numerical Simulation}
In this section, we conduct numerical simulations to evaluate the performance of the proposed control framework. Section~\ref{subsec: Simulation Scenario} displays the studied scenarios including mixed-autonomy platoon~\cite{zhou2024enhancing}, drone formation control~\cite{riehl2022string}, and microgrid~\cite{silva2021string}. Section~\ref{subsec: Training details} presents the training procedure and simulation visualizations. Section~\ref{subsec: verification results} introduces the verification effectiveness of the proposed method. Section~\ref{subsec: controller results} presents the simulation results. 

\subsection{Experimental Setup}
\label{subsec: Simulation Scenario}
\noindent\textbf{Mixed-autonomy platoon.} We consider a mixed-autonomy platoon with CAVs $\Omega_{\mathcal{C}}$ and HDVs $\Omega_{\mathcal{H}}$, where $n=|\Omega_{\mathcal{C}}\cup\Omega_{\mathcal{H}}|$ and $m=|\Omega_{\mathcal{C}}|$ denote the total number of vehicles and CAVs, respectively. The discrete-time longitudinal motion of vehicle $i\in\Omega_{C}\cup\Omega_{\mathcal{H}}$ with sampling period $T>0$ is  
\begin{align}
&s_{i,k+1} = s_{i,k} + T[v_{i-1,k} - v_{i,k}], \\
&v_{i,k+1} = v_{i,k} + T 
\begin{cases}
u_{i,k}, & i \in \Omega_\mathcal{C},\\
\mathbb{F}_i(s_{i,k},v_{i,k},v_{i-1,k}), & i \in \Omega_\mathcal{H},
\end{cases}
\end{align}
where $s_{i,k}$ and $v_{i,k}$ are spacing and velocity at step $k$.  
CAV control $u_{i,k}$ follows an RL policy~\cite{zhou2024enhancing}, while HDVs follow an unknown car-following model like the Full Velocity Difference (FVD) model~\cite{jiang2001full}.

\noindent \textbf{Drone formation.}
We consider a leader–follower drone formation in $\mathbb{R}^3$ with one leader and $N_f$ followers.  
Let $p_\ell,v_\ell,u_\ell \in \mathbb{R}^3$ denote the leader's position, velocity, and acceleration, respectively, and $p_i,v_i,u_i$ the corresponding states for follower $i$.  
The leader follows a predefined trajectory:
\begin{align}
p_{\ell,k+1} &= p_{\ell,k} + T\, v_{\ell,k},\\
v_{\ell,k+1} &= v_{\ell,k} + T\, u_{\ell,k}.
\end{align}
Each follower follows:
\begin{equation}
x_{i,k+1} = \mathrm{F}_i\big(x_{i,k},u_{i,k},x^r_{i,k}\big),
\end{equation}
with $x_{i,k}=[p_{i,k},v_{i,k}]^\top$ and $x^r_{i,k}$ the reference state (leader or preceding drone). The true dynamic is unknown $\mathrm{F}_i$, and we estimate it from data. The follower control input is generated by a neural policy trained via supervised learning to imitate a known controller (e.g., optimal LQR feedback).

\noindent \textbf{Microgrid.} We consider $n$ voltage-source inverters indexed by $\mathcal{N}=\{1,\dots,n\}$, connected in a radial topology with neighbors $\mathcal{E}_i$.  
Each node $i$ has phase angle $\delta_{i,k}$, voltage magnitude $U_{i,k}>0$, and frequency $\omega_{i,k}$. The state is represented as $x_{i,k} = (\delta_{i,k},\omega_{i,k},\xi_{i,k})$.  
The active power at $i$ is
\begin{equation}
P_{i,k} = P_{L,i} + \sum_{j\in\mathcal{E}_i} \alpha_{ij} \sin(\delta_{i,k}-\delta_{j,k}),\quad
\alpha_{ij} = |B_{ij}|U_{i,k}U_{j,k},
\end{equation}
where $P_{L,i}$ is the load demand.

The closed-loop dynamics are
\begin{align}
&\delta_{i,k+1} = \delta_{i,k} + T\omega_{i,k},\\
&\omega_{i,k+1} = \omega_{i,k} + \frac{T}{\tau_i}
\big[ -(\omega_{i,k}-\omega^*)\notag\\
& - \eta_i(P_{i,k}-P_i^*) + \xi_{i,k} \big],\\
&\xi_{i,k+1} = \xi_{i,k} + Tu_{i,k}.
\end{align}
where $\tau_i>0$ denotes the first-order time constant of the frequency loop, and $\eta_i>0$ is the active-power droop gain, $P_i^*$ is the power setpoint, $\omega^*$ is the nominal frequency.
The secondary controller updates with a neural policy:
\begin{equation}
u_{i,k} = \pi_{\theta_i}\big(x_{i,k},\{x_{j,k}\}_{j\in\mathcal{E}_i}\big),
\end{equation}

\subsection{Training and Simulation Demonstration}
\label{subsec: Training details}
\begin{figure*}[htp]
    \centering
    \begin{subfigure}[t]{0.32\textwidth}
        \centering
        \includegraphics[width=\linewidth]{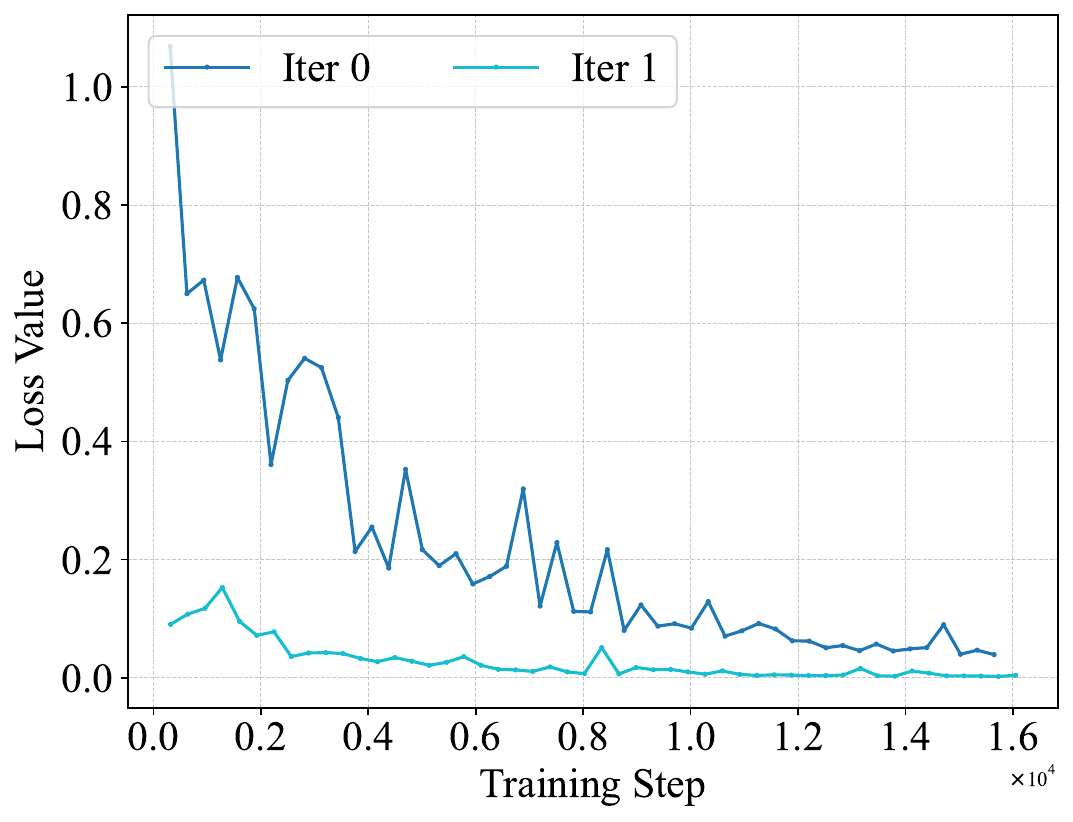}
        \caption{Platoon}
        \label{fig:train_loss_platoon}
    \end{subfigure}
    \begin{subfigure}[t]{0.31\textwidth}
        \centering
        \includegraphics[width=\linewidth]{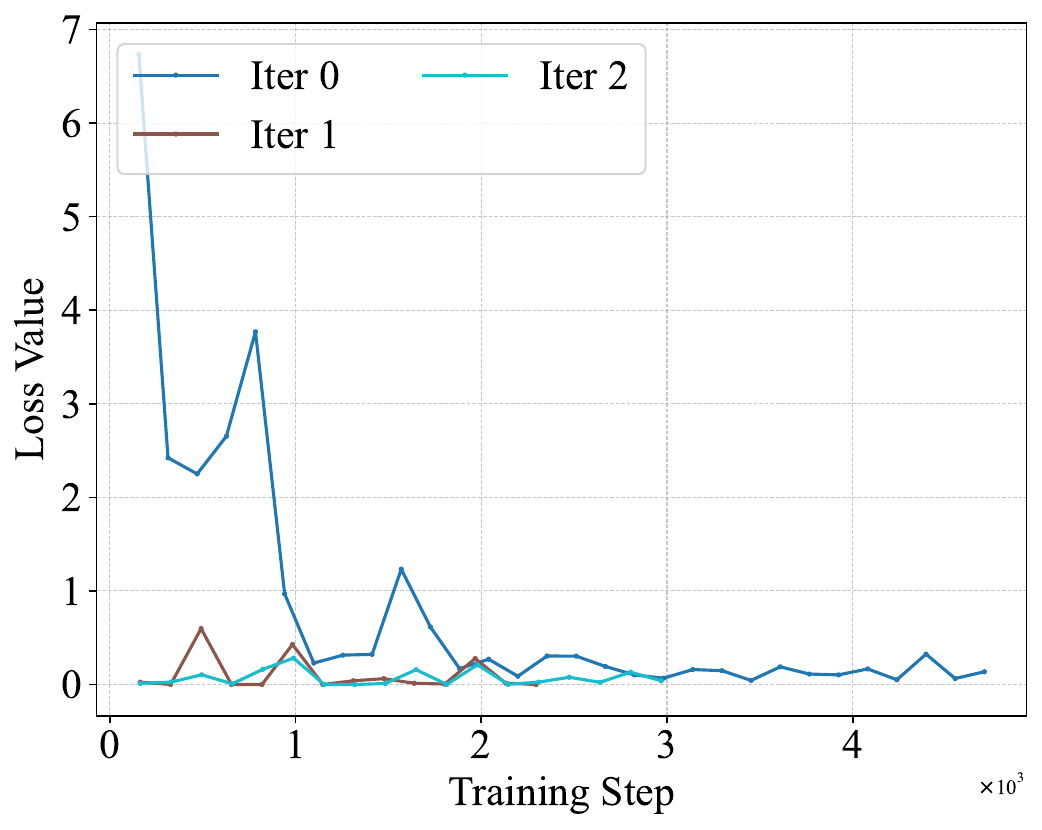}
        \caption{Drones formation}
        \label{fig:train_loss_uavs}
    \end{subfigure}
    \begin{subfigure}[t]{0.325\textwidth}
        \centering
        \includegraphics[width=\linewidth]{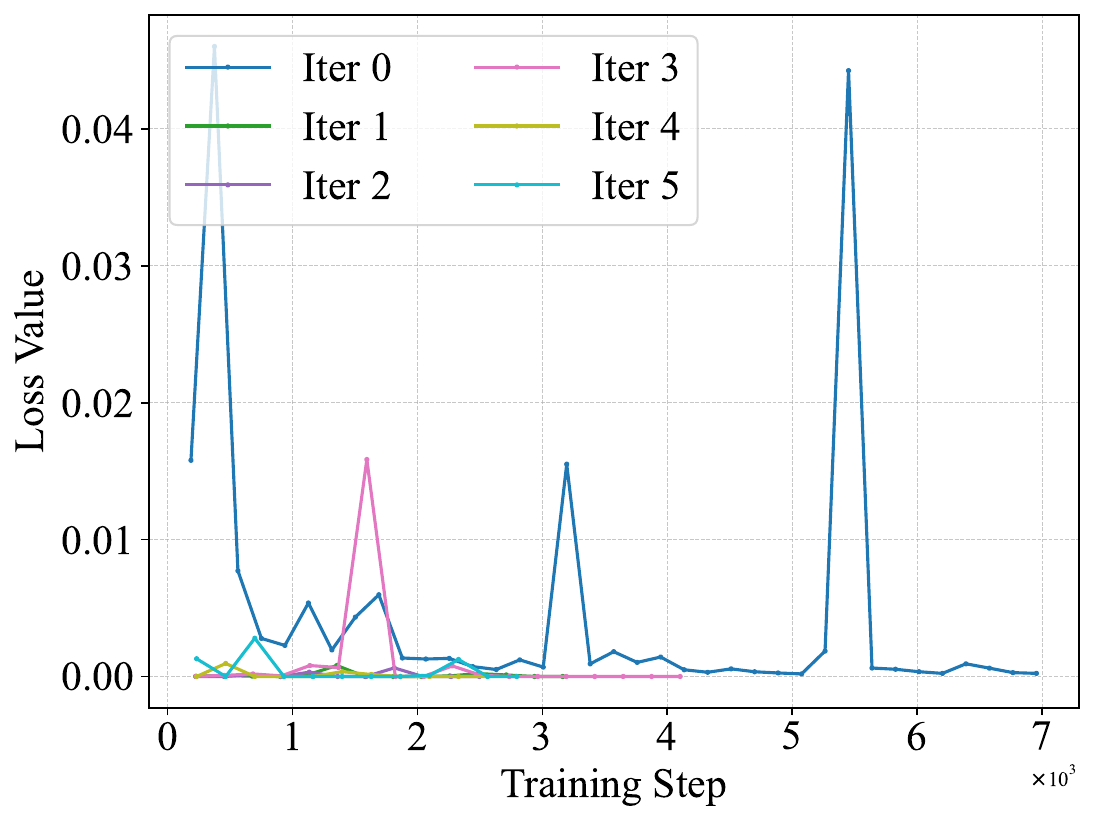}
        \caption{Microgrids}
        \label{fig:train_loss_smart_grids}
    \end{subfigure}
    \caption{Training loss curves in different environments.}
    \label{fig:train_loss_all}
\end{figure*}

 \noindent\textbf{Training Details.} Each CEGIS loop is trained for a maximum of 100 epochs, with a maximum of 100 iterations. The learning rate is initialized at 0.001 and follows a decay schedule. The initial dataset comprises 30,000 state pairs, with 80\% used for training and 20\% for validation, and a batch size of 32. To augment the counter-example set, Gaussian noise is applied to each original counter-example, generating 20 variants per instance. Both the Lyapunov function and the controller are implemented as three-layer fully connected neural networks, each with a hidden dimension of 64. The activation function used is ReLU for all hidden layers. All runtime-related experiments were conducted three times, and the results are reported as the mean and standard deviation to capture both central tendency and variability. This repetition helps assess the consistency and statistical reliability of the performance measurements. 

 Fig.~\ref{fig:train_loss_all} shows the training loss trajectories in three representative environments: (a) Platoon, (b) Drones formation, and (c) Microgrids. Each iteration corresponds to a distinct training loop triggered by the addition of a new counterexample. Specifically, the platoon scenario contains 2 iterations, the drone scenario includes 3 iterations, and the microgrid scenario involves 6 iterations. In each environment, a new iteration begins when counterexamples are introduced, leading to an initial increase in loss as the policy is updated to address the new scenario. Over time, the loss decreases and stabilizes, indicating that the policy gradually adapts and learns to handle the updated data.

 \begin{figure}[h]
    \centering
    \begin{subfigure}[t]{0.23\textwidth}
        \includegraphics[width=\linewidth]{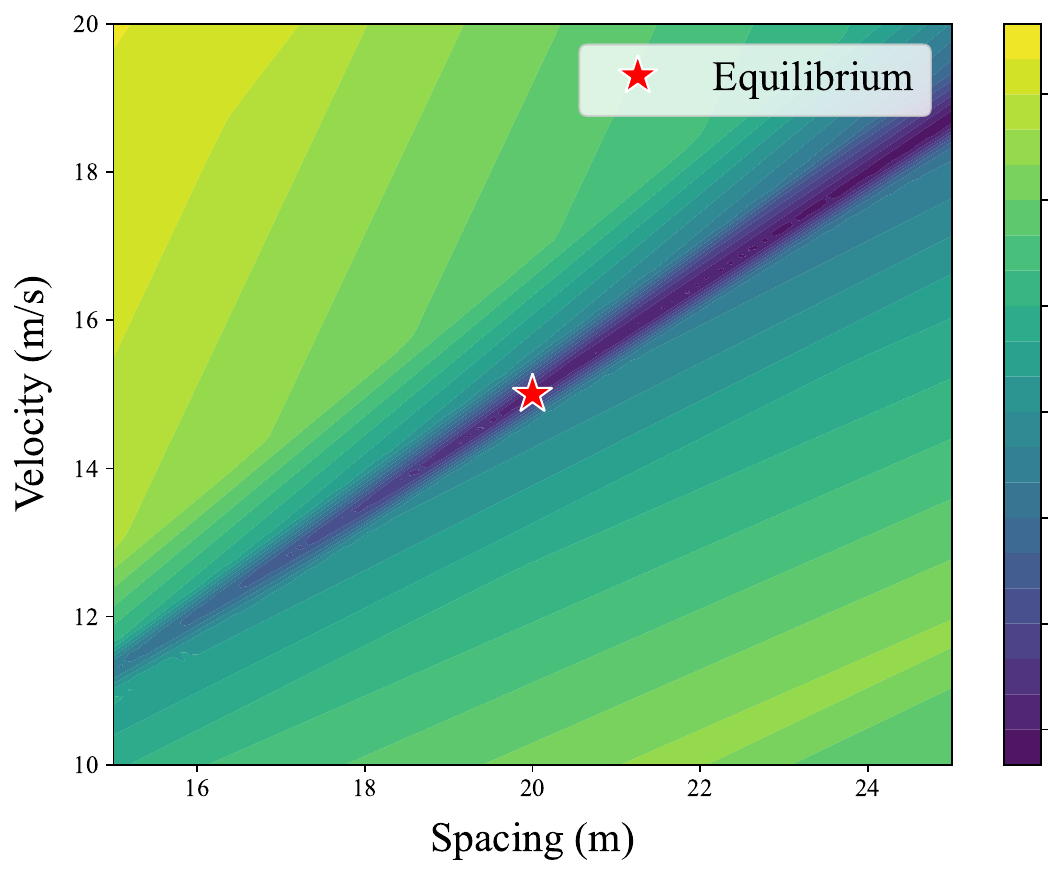}
        \caption{CAV Lyapunov contour}
        \label{fig:lyap_cav}
    \end{subfigure}
    \hfill
    \begin{subfigure}[t]{0.23\textwidth}
        \centering
        \includegraphics[width=\linewidth]{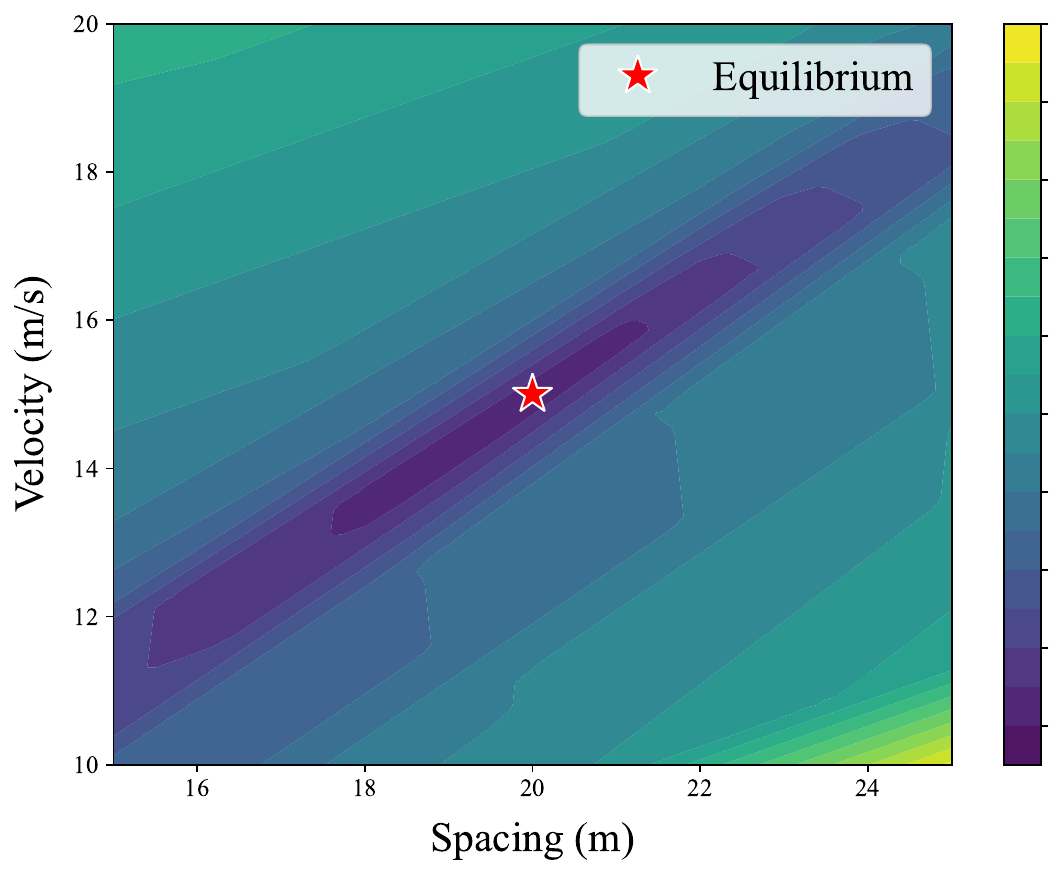}
        \caption{HDV Lyapunov contour}
        \label{fig:lyap_hdv}
    \end{subfigure}
    \caption{Learned Lyapunov contours for CAV and HDV systems.}
    \label{fig:lyap_contour_platoon}
\end{figure}

\begin{figure}[h]
    \centering
    \begin{subfigure}[t]{0.23\textwidth}
        \centering
        \includegraphics[width=\linewidth]{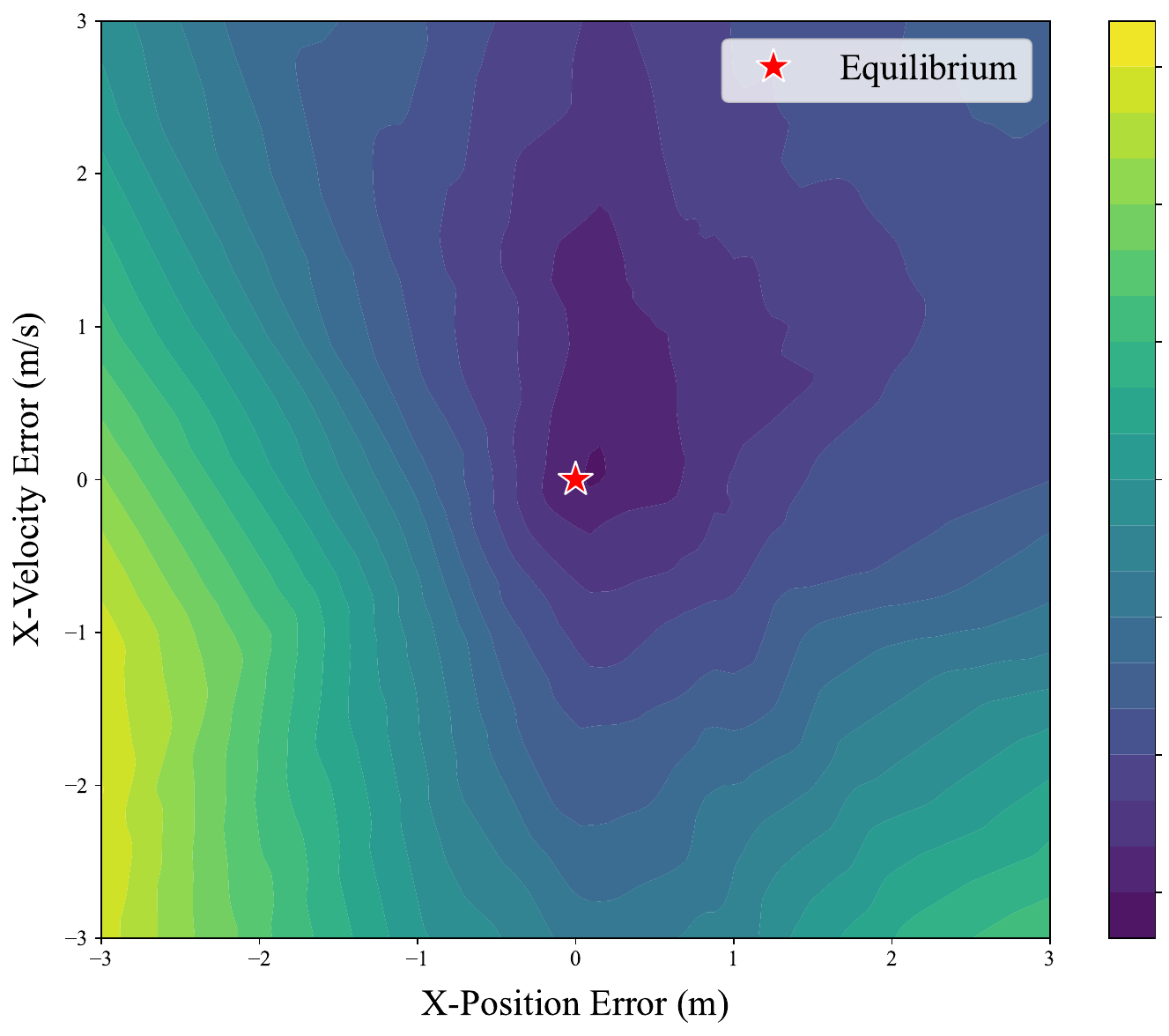}
        \caption{Drone 1}
        \label{fig:uav1_lyapunov}
    \end{subfigure}
    \hfill
    \begin{subfigure}[t]{0.23\textwidth}
        \centering
        \includegraphics[width=\linewidth]{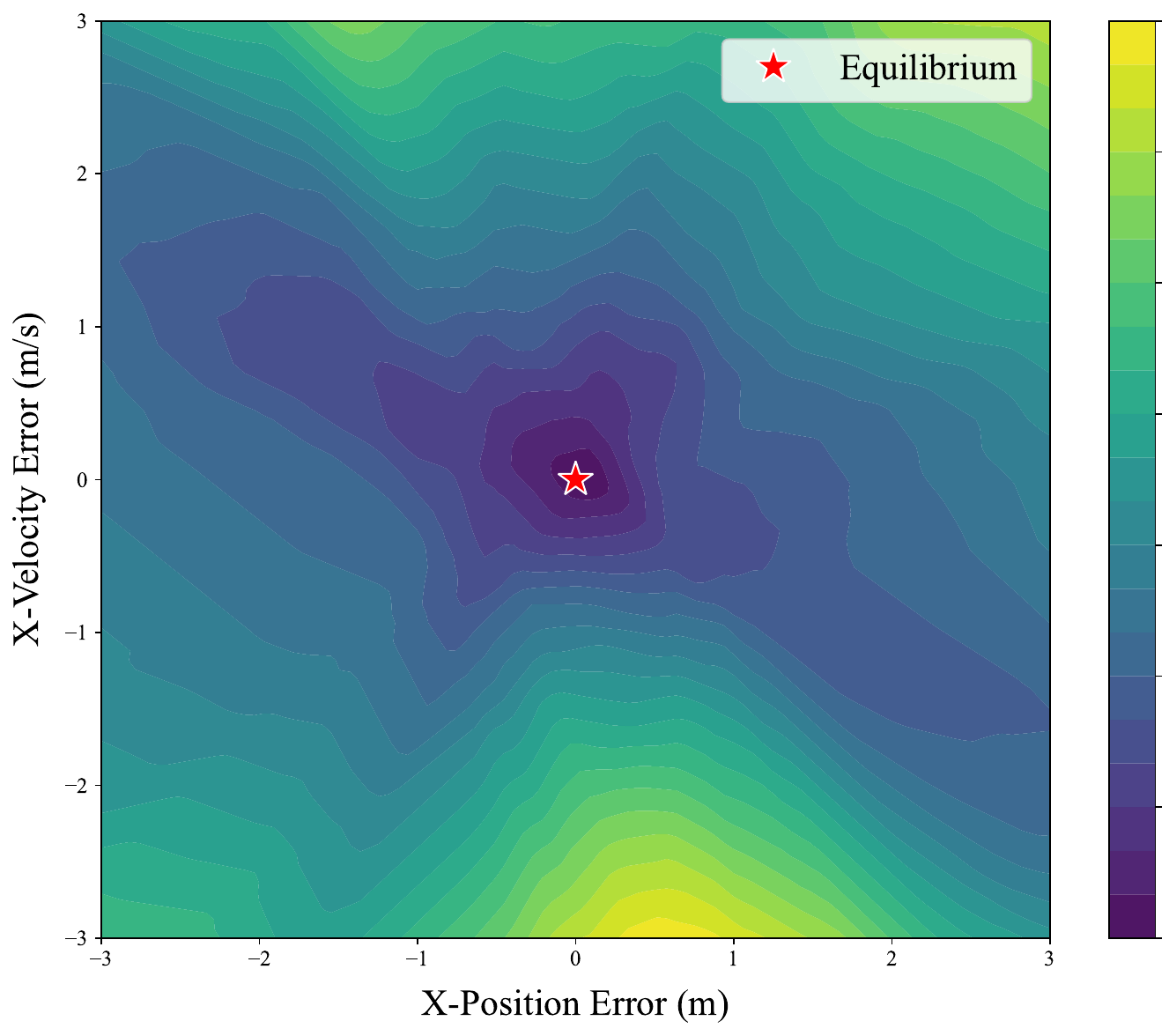}
        \caption{Drone 2}
        \label{fig:uav2_lyapunov}
    \end{subfigure}
    \caption{2D Lyapunov function visualizations for two Drones in the Drones formation control.}
    \label{fig:uav_lyapunov}
\end{figure}

\begin{figure}[h]
    \centering
    \begin{subfigure}[t]{0.23\textwidth}
        \centering
        \includegraphics[width=\linewidth]{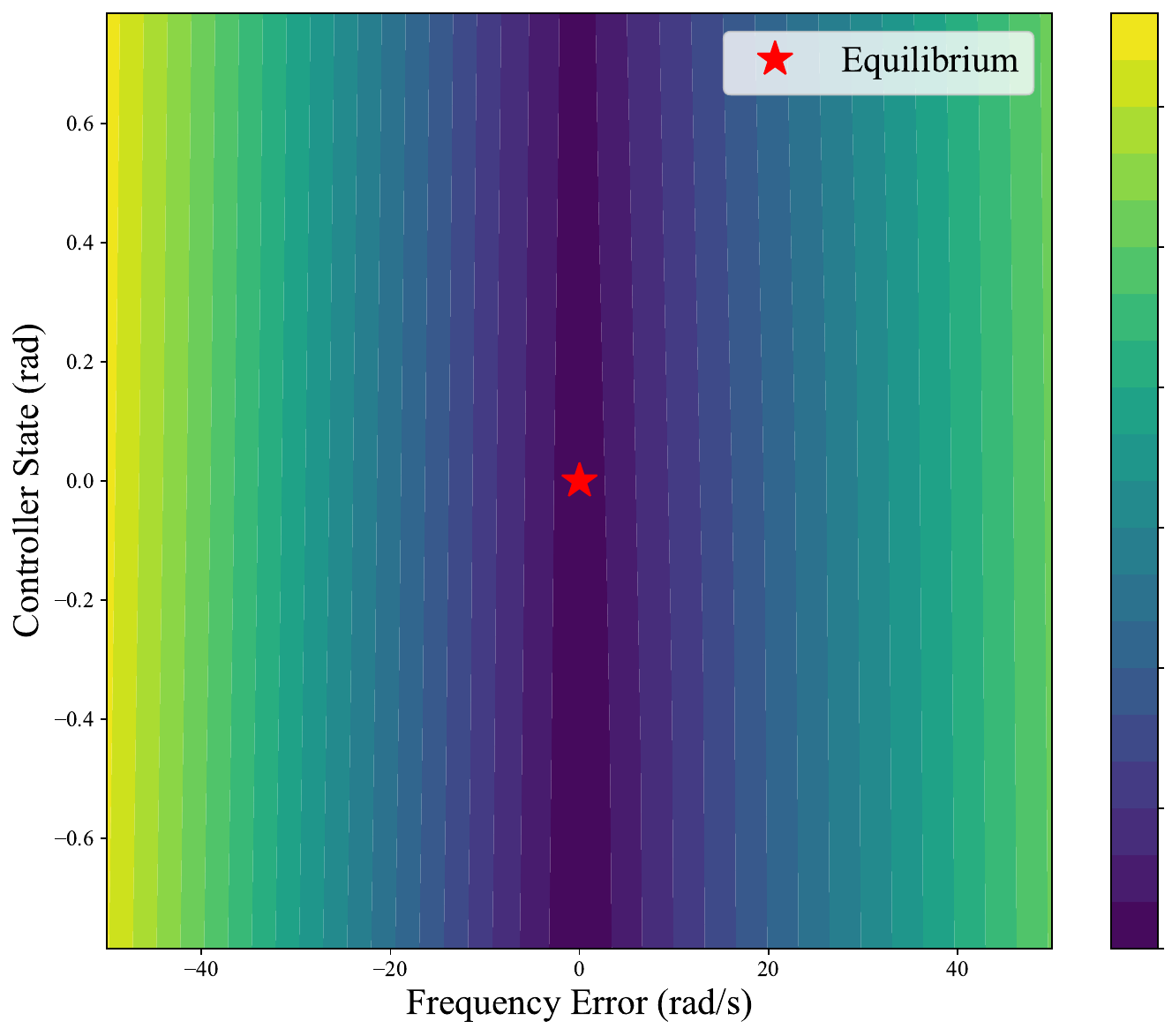}
        \caption{Inverter 1}
        \label{fig:lyapunov_inverter1}
    \end{subfigure}
    \begin{subfigure}[t]{0.23\textwidth}
        \centering
        \includegraphics[width=\linewidth]{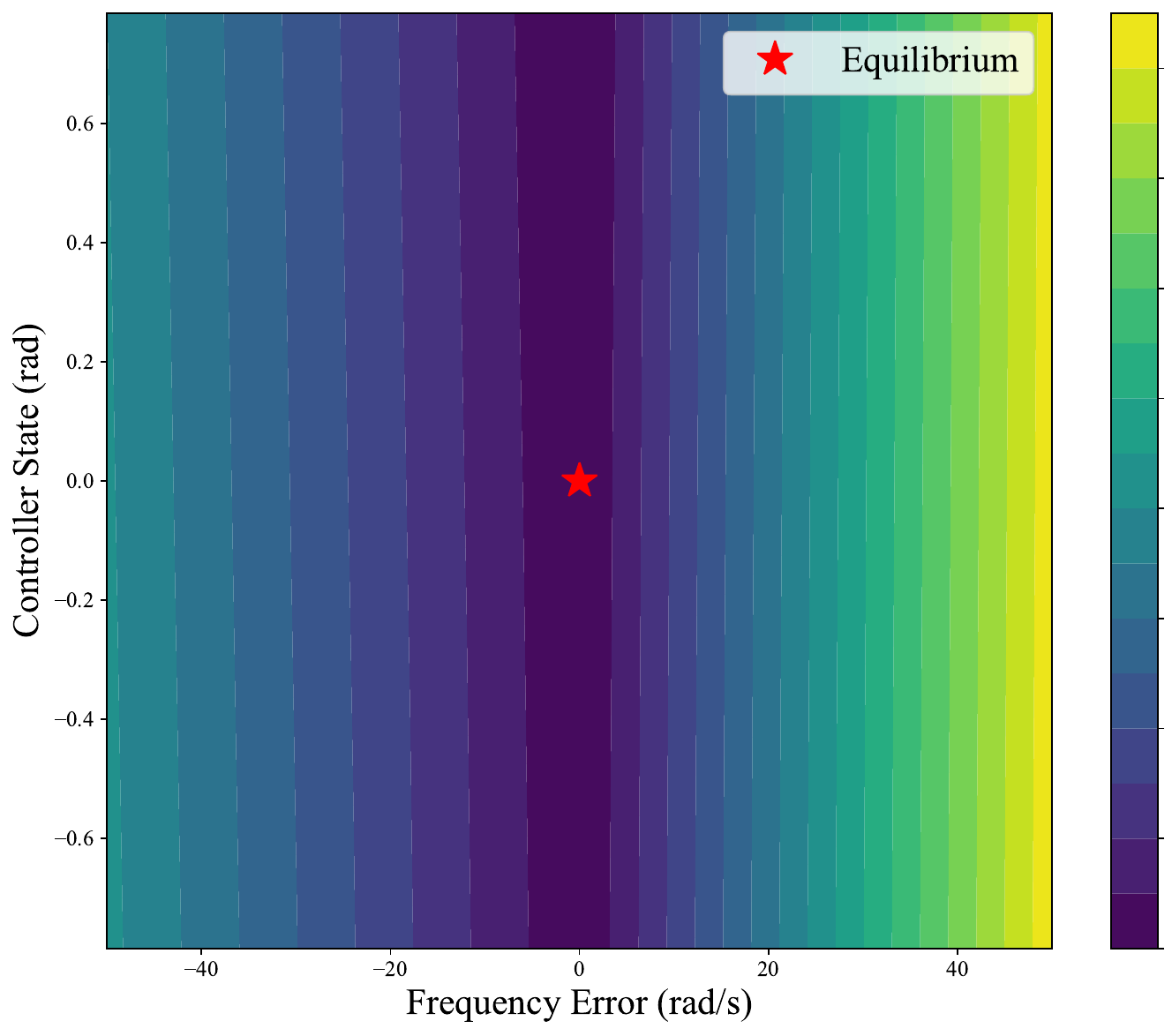}
        \caption{Inverter 2}
        \label{fig:lyapunov_inverter2}
    \end{subfigure}
    \caption{2D Lyapunov function visualizations for two inverters in the microgrid system.}
    \label{fig:lyapunov_microgrid}
\end{figure}

\noindent\textbf{Lyapunov Visualization.} Figs.~\ref{fig:lyap_contour_platoon},~\ref{fig:uav_lyapunov}, and \ref{fig:lyapunov_microgrid} illustrate 2D visualizations of the learned and verified certificates across multiple domains, including mixed-autonomy platoons, drone formation control, and microgrid inverters. In each subplot, the equilibrium point is marked by a red star, and the contour plots represent the value of the Lyapunov function over selected state slices. These visualizations qualitatively demonstrate the key property of Lyapunov functions: values increasing with distance from the equilibrium. The contours reflect that the learned functions generally capture the local stability structure around the equilibrium across different dynamics.

\subsection{Verification Performance}
\label{subsec: verification results}

We carry out two verification studies.  

(1) The \emph{large‑scale test} is used to evaluate the scalability of our framework to large-scale interconnected systems by exploiting equivalent node structure according to Theorems~\ref{thm:Certifiable Equivalence} and Corollary~\ref{col: eq_nodes}. We compare full retraining and verification (\emph{Full R.}) with the reduced‑verification nodes determined by Alg.~\ref{alg:mvn-reduction} (denoted \emph{RedVer}).  

(2) The \emph{additive‑topology test} is used to evaluate the value of decomposability according to Def.~\ref{dfn:additiveness} and Corollary~\ref{col:augmentedSystem}. A new sub‑network of various sizes is attached to a pre‑trained system. We compare full retraining and verification (\emph{Full R.}) with controller reuse plus local training (\emph{AddReuse}).  

For every simulation run, we record the average retraining time per epoch (RT), the total verification time (VT), and their sum (Tot). \emph{TO} indicates time out, i.e., failed to complete within 4 hours of training and verification.

Table~\ref{tab:large-scale} summarizes the results of the large-scale test, where the number of agents in each system is progressively increased. The proposed method RedVer significantly reduces both retraining time (RT) and verification time (VT) compared to full retraining (Full R). For example, in the mixed-autonomy platoon with 50 agents, RedVer completes verification with an average time of 1108 seconds, while Full R results in a timeout. Similar improvements are observed in the drone formation and microgrid scenarios, demonstrating the scalability of the RedVer strategy.

\begin{table}[htp]
\centering
\caption{Large‑scale test.}
\label{tab:large-scale}
\setlength{\tabcolsep}{3pt}
\begin{tabular}{l c c c c c}
\toprule
Scenario & \# Agent & Method & RT (s) & VT (s) & Tot (s) \\ \midrule
\multirow{4}{*}{\makecell[l]{Mixed-\\autonomy\\platoon}} 
      & 10 & Full R.  &1633$\pm$58  & 138$\pm$53& 1773$\pm$109\\ 
      & 10 & RedVer &1066$\pm$10&41$\pm$18 &1108$\pm$25 \\ 
      & 50 & Full R. & - & - & TO\\ 
      & 50 & RedVer &1072$\pm$20 & 47$\pm$19 & 1108$\pm$26\\ \midrule
\multirow{4}{*}{\makecell[l]{Drones\\formation}}    
      & 6  & Full R.  &- &- &TO \\ 
      & 6  & RedVer &411$\pm$162 &785$\pm$152 & 1200$\pm$290\\ 
      & 20 & Full R.  &- & -&TO \\ 
      & 20 & RedVer &414$\pm$191&791$\pm$149 &1210$\pm$336\\ \midrule
\multirow{4}{*}{Microgrid} 
      & 10  & Full R.  &- &- &TO \\ 
      & 10  & RedVer & 392$\pm$102 &521$\pm$180 & 920$\pm$264\\ 
      & 50 & Full R.  &- &- &TO \\ 
      & 50 & RedVer &436$\pm$106 &614$\pm$23 & 1051$\pm$119\\ \bottomrule
\end{tabular}
\end{table}

\begin{table}[ht]
  \centering
  \caption{Additive-topology test.}
  \label{tab:additive}
  \small
  \setlength{\tabcolsep}{4pt}
  \begin{tabular}{l cc cc}
    \toprule
    & \multicolumn{2}{c}{Mixed-autonomy platoon} & \multicolumn{2}{c}{Drones formation}\\
    \cmidrule(lr){2-3}\cmidrule(lr){4-5}
    Metric & Full R. & AddReuse & Full R. & AddReuse \\ \midrule
    Orig. Nodes & 5 & 5 & 3 & 3 \\
    New Nodes      & 5 & 5 & 1 & 1 \\
    RT (s)            & 1632$\pm$131 & 1592$\pm$32  & 200$\pm$53 & 251$\pm$48 \\
    VT (s)            & 236$\pm$97  & 61$\pm$32   & 505$\pm$18 & 108$\pm$37 \\
    Tot (s)           & 1868$\pm$227 & 1653$\pm$30 & 706$\pm$71 & 360$\pm$69 \\ 
    \bottomrule
  \end{tabular}
\end{table}
Table~\ref{tab:additive} shows the results for the additive-topology test, in which a new sub-network is added to a pre-trained system. The AddReuse method, which reuses existing controllers with localized training, achieves substantial savings in runtime compared to Full R. In particular, for the mixed-autonomy platoon task, AddReuse reduces verification time from 236 seconds to 61 seconds, effectively reducing the total verification effort.

These results confirm the effectiveness of the proposed verification strategies in both scalability and decomposability settings, which is achieved by reusing certain local certificates.

\subsection{Controller Performance}
\label{subsec: controller results}
The string stability condition is commonly characterized by the boundedness of the velocity error gain between neighboring agents, expressed as:
\begin{align}
\max_i\frac{\|v_{i}\|_{\mathcal{L}_2}}{\|v_{i-1}\|_{\mathcal{L}_2}} \leq 1 
\end{align}
which ensures that velocity perturbations do not grow downstream.

To empirically validate this, we applied sinusoidal disturbances with varying frequencies and amplitudes and measured the resulting maximum error gain.

\begin{table}[htp]
  \centering
  \caption{Maximum Error Gain.}
  \label{tab:error-gain}
  \small
  \setlength{\tabcolsep}{4pt}
  \begin{tabular}{lcc}
    \toprule
    & 1/15 Hz, 4 m/s & 1/15 Hz, 7 m/s \\ \midrule
    Original $\pi$              & 1.0011 & 4.6056 \\
    Compositional ISS~\cite{zhang2023compositional}      & 0.9954 & 1.1157 \\
    Proposed Method             & \textbf{0.9931} & \textbf{0.9932} \\ 
    \bottomrule
  \end{tabular}
\end{table}

As shown in Table~\ref{tab:error-gain}, the original policy leads to amplification (gain $>$ 1), and under large, low-frequency disturbances, the amplification becomes severe (e.g., 4.6056). While the ISS baseline maintains robustness in moderate conditions, it does not guarantee string stability and fails under stronger perturbations. In contrast, our proposed method consistently maintains gains below 1.0, successfully enforcing string stability.

\begin{figure}[htp]
    \centering
    \begin{subfigure}[t]{0.45\textwidth}
        \centering
        \includegraphics[width=\linewidth]{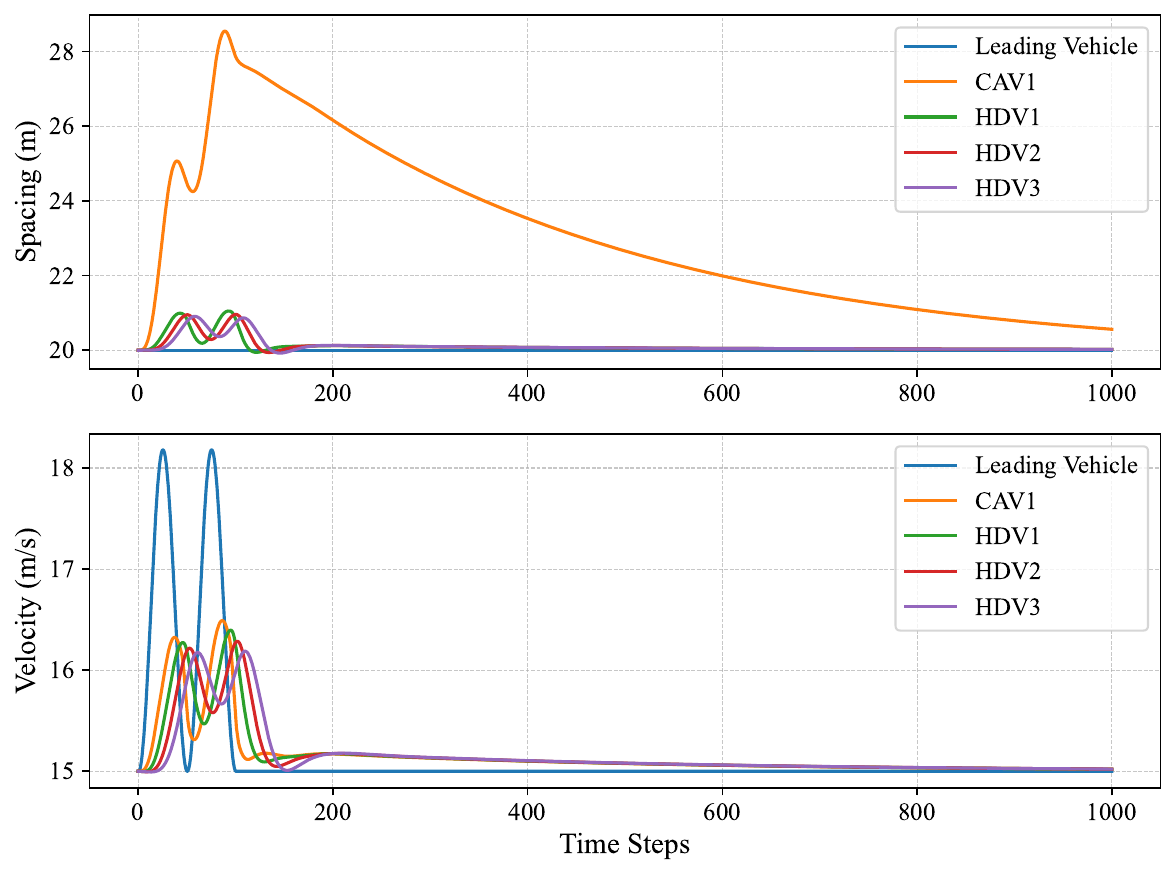}
        \caption{Mixed-autonomy platoon trajectories}
        \label{fig:trajectory_platoon}
    \end{subfigure}
    \begin{subfigure}[t]{0.45\textwidth}
        \centering
        \includegraphics[width=\linewidth]{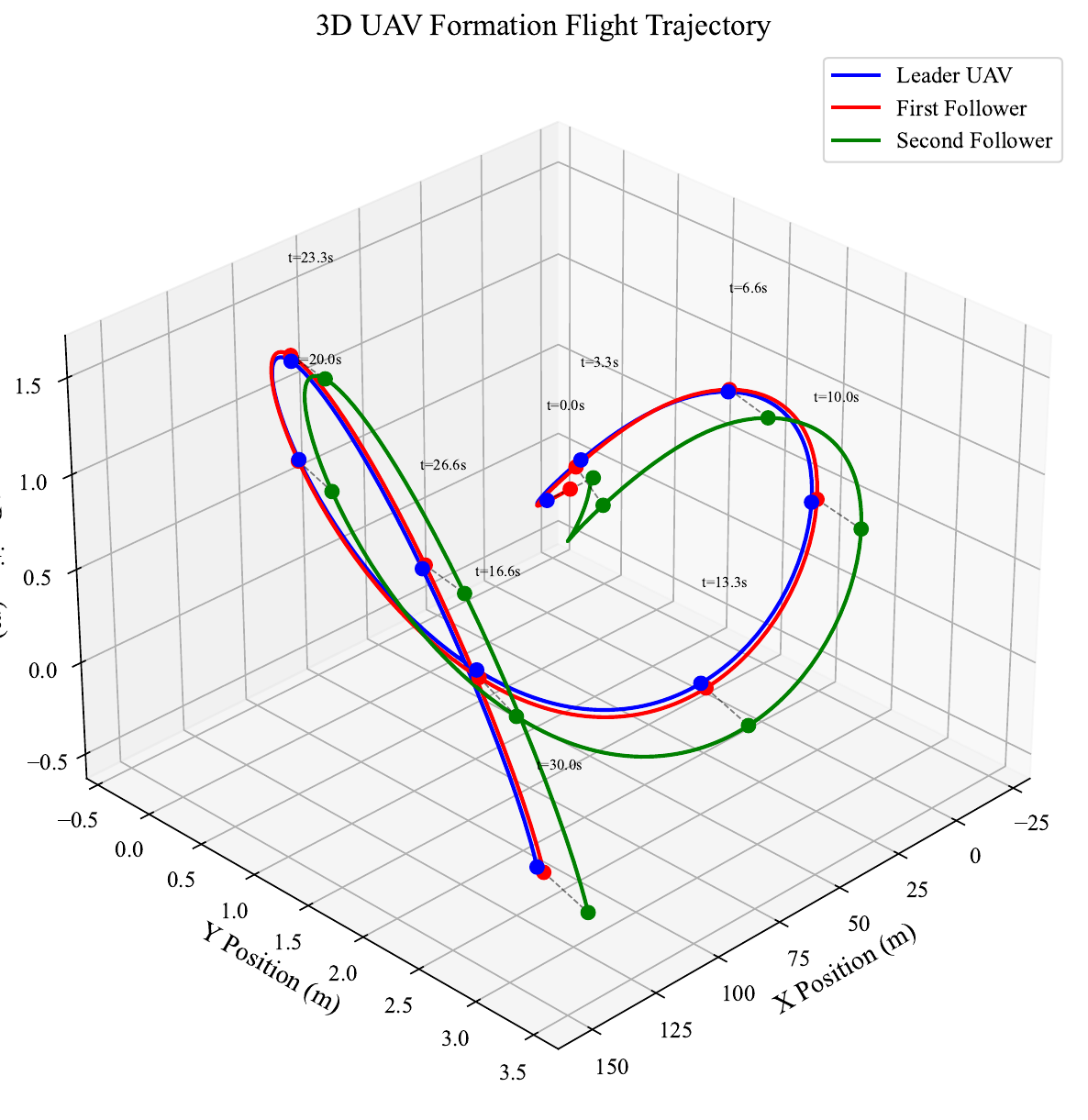}
        \caption{Drone formation trajectories}
        \label{fig:drones_platoon}
    \end{subfigure}
    \begin{subfigure}[t]{0.45\textwidth}
        \centering
        \includegraphics[width=\linewidth]{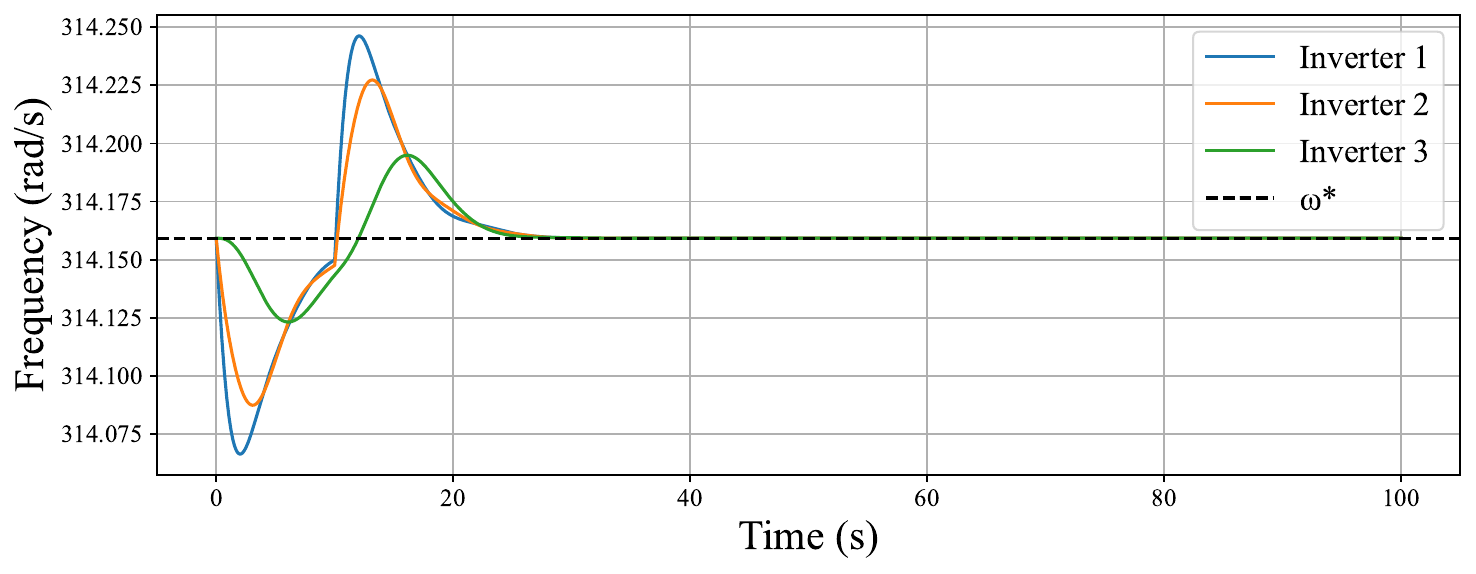}
        \vspace{0.5em}
        \includegraphics[width=\linewidth]{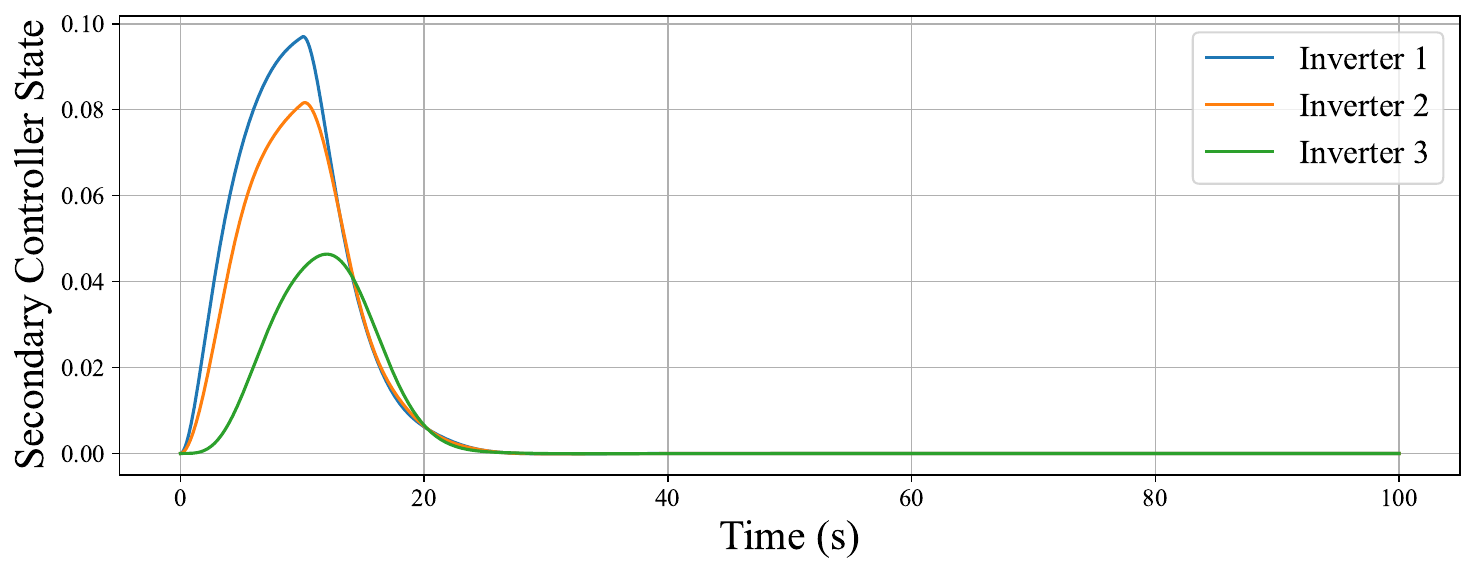}
        \caption{Frequencies and secondary controller states for the microgrid}
        \label{fig:microgrid_trajectories}
    \end{subfigure}

    \caption{Trajectory visualizations in three environments: 
    (a) mixed-autonomy platoon, 
    (b) drone formation, 
    (c) microgrid.}
    \label{fig:all_trajectories}
\end{figure}

Moreover, we visualize the response of the verified controllers to the disturbance in three scenarios. Fig.~\ref{fig:trajectory_platoon} to~\ref{fig:microgrid_trajectories} present the system responses when the leading agent in each scenario experiences a temporary disturbance. Specifically, a perturbation is applied to the front CAV in the platoon, the leader Drone in formation flight, and the primary reference of inverter 1 in the microgrid. Despite these perturbations, the remaining agents exhibit smooth recovery and converge to their desired states, indicating that the proposed controllers maintain string stability and coordination under disturbance.

\section{Conclusion}
In this paper, we presented a scalable synthesis and verification framework that combines discrete-time scalable input-to-state stability certificates with neural-network verification to provide formal string-stability guarantees for large-scale interconnected systems. We developed a robust verification procedure that explicitly bounds the mismatch between the true dynamics and their neural surrogates. We further established theory and algorithms that scale training and verification to large-scale interconnected systems, and extended the framework to systems with external inputs, enabling simultaneous synthesis and verification of neural certificates and controllers. Numerical studies on mixed-autonomy platoons, drone formations, and microgrids demonstrate that the approach certifies string stability and efficiently produces verified controllers at scale.

\bibliographystyle{unsrt}       
\bibliography{autosam}           

\begin{thebibliography}{10}

\bibitem{haber2014subspace}
Aleksandar Haber and Michel Verhaegen.
\newblock Subspace identification of large-scale interconnected systems.
\newblock {\em IEEE Transactions on Automatic Control}, 59(10):2754--2759, 2014.

\bibitem{lyu2022small}
Ziliang Lyu, Xiangru Xu, and Yiguang Hong.
\newblock Small-gain theorem for safety verification of interconnected systems.
\newblock {\em Automatica}, 139:110178, 2022.

\bibitem{zhang2025privacy}
Yuchen Zhang, Bo~Chen, Jianzheng Wang, and Li~Yu.
\newblock Privacy-preserving distributed estimation for interconnected dynamic systems.
\newblock {\em Automatica}, 177:112277, 2025.

\bibitem{janssen2024modular}
Lars~AL Janssen, Bart Besselink, Rob~HB Fey, and Nathan van~de Wouw.
\newblock Modular model reduction of interconnected systems: A robust performance analysis perspective.
\newblock {\em Automatica}, 160:111423, 2024.

\bibitem{silva2021string}
Guilherme~F Silva, Alejandro Donaire, Maria~M Seron, Aaron McFadyen, and Jason Ford.
\newblock String stability in microgrids using frequency controlled inverter chains.
\newblock {\em IEEE Control Systems Letters}, 6:1484--1489, 2021.

\bibitem{wang2022optimal}
Yan Wang, Rong Su, and Bohui Wang.
\newblock Optimal control of interconnected systems with time-correlated noises: Application to vehicle platoon.
\newblock {\em Automatica}, 137:110018, 2022.

\bibitem{zhou2024parameter}
Jingyuan Zhou and Kaidi Yang.
\newblock A parameter privacy-preserving strategy for mixed-autonomy platoon control.
\newblock {\em Transportation Research Part C: Emerging Technologies}, 169:104885, 2024.

\bibitem{zhu2002model}
Guang-Yan Zhu and Michael~A Henson.
\newblock Model predictive control of interconnected linear and nonlinear processes.
\newblock {\em Industrial \& engineering chemistry research}, 41(4):801--816, 2002.

\bibitem{feng2019string}
Shuo Feng, Yi~Zhang, Shengbo~Eben Li, Zhong Cao, Henry~X Liu, and Li~Li.
\newblock String stability for vehicular platoon control: Definitions and analysis methods.
\newblock {\em Annual Reviews in Control}, 47:81--97, 2019.

\bibitem{wang2021leading}
Jiawei Wang, Yang Zheng, Chaoyi Chen, Qing Xu, and Keqiang Li.
\newblock Leading cruise control in mixed traffic flow: System modeling, controllability, and string stability.
\newblock {\em IEEE Transactions on Intelligent Transportation Systems}, 23(8):12861--12876, 2021.

\bibitem{gratzer2022string}
Alexander~L Gratzer, Sebastian Thormann, Alexander Schirrer, and Stefan Jakubek.
\newblock String stable and collision-safe model predictive platoon control.
\newblock {\em IEEE Transactions on Intelligent Transportation Systems}, 23(10):19358--19373, 2022.

\bibitem{aboudonia2025adaptive}
Ahmed Aboudonia and John Lygeros.
\newblock Adaptive learning-based model predictive control for uncertain interconnected systems: A set membership identification approach.
\newblock {\em Automatica}, 171:111943, 2025.

\bibitem{guo2016distributed}
Xianggui Guo, Jianliang Wang, Fang Liao, and Rodney Swee~Huat Teo.
\newblock Distributed adaptive integrated-sliding-mode controller synthesis for string stability of vehicle platoons.
\newblock {\em IEEE Transactions on Intelligent Transportation Systems}, 17(9):2419--2429, 2016.

\bibitem{cheng2019end}
Richard Cheng, G{\'a}bor Orosz, Richard~M Murray, and Joel~W Burdick.
\newblock End-to-end safe reinforcement learning through barrier functions for safety-critical continuous control tasks.
\newblock In {\em Proceedings of the AAAI conference on artificial intelligence}, volume~33, pages 3387--3395, 2019.

\bibitem{li2021reinforcement}
Meng Li, Zehong Cao, and Zhibin Li.
\newblock A reinforcement learning-based vehicle platoon control strategy for reducing energy consumption in traffic oscillations.
\newblock {\em IEEE Transactions on Neural Networks and Learning Systems}, 32(12):5309--5322, 2021.

\bibitem{zhou2024enhancing}
Jingyuan Zhou, Longhao Yan, and Kaidi Yang.
\newblock Enhancing system-level safety in mixed-autonomy platoon via safe reinforcement learning.
\newblock {\em IEEE Transactions on Intelligent Vehicles}, pages 1--13, 2024.

\bibitem{zhang2024string}
Xiaohui Zhang, Jie Sun, Zuduo Zheng, and Jian Sun.
\newblock On the string stability of neural network-based car-following models: A generic analysis framework.
\newblock {\em Transportation research part C: emerging technologies}, 160:104525, 2024.

\bibitem{dai2021lyapunov}
Hongkai Dai, Benoit Landry, Lujie Yang, Marco Pavone, and Russ Tedrake.
\newblock Lyapunov-stable neural-network control.
\newblock {\em Robotics: Science and Systems}, 2021.

\bibitem{yanglyapunov}
Lujie Yang, Hongkai Dai, Zhouxing Shi, Cho-Jui Hsieh, Russ Tedrake, and Huan Zhang.
\newblock Lyapunov-stable neural control for state and output feedback: A novel formulation.
\newblock In {\em Forty-first International Conference on Machine Learning}, 2024.

\bibitem{mandal2024safe}
Udayan Mandal, Guy Amir, Haoze Wu, Ieva Daukantas, Fletcher~Lee Newell, Umberto Ravaioli, Baoluo Meng, Michael Durling, Kerianne Hobbs, Milan Ganai, et~al.
\newblock Safe and reliable training of learning-based aerospace controllers.
\newblock In {\em 2024 AIAA DATC/IEEE 43rd Digital Avionics Systems Conference (DASC)}, pages 1--10. IEEE, 2024.

\bibitem{mandal2024formally}
Udayan Mandal, Guy Amir, Haoze Wu, Ieva Daukantas, Fletcher~Lee Newell, Umberto~J Ravaioli, Baoluo Meng, Michael Durling, Milan Ganai, Tobey Shim, et~al.
\newblock Formally verifying deep reinforcement learning controllers with lyapunov barrier certificates.
\newblock {\em arXiv preprint arXiv:2405.14058}, 2024.

\bibitem{zhang2023compositional}
Songyuan Zhang, Yumeng Xiu, Guannan Qu, and Chuchu Fan.
\newblock Compositional neural certificates for networked dynamical systems.
\newblock In {\em Learning for Dynamics and Control Conference}, pages 272--285. PMLR, 2023.

\bibitem{silva2024scalable}
Guilherme~Fr{\'o}es Silva, Alejandro Donaire, Richard Middleton, Aaron McFadyen, and Jason Ford.
\newblock Scalable input-to-state stability of nonlinear interconnected systems.
\newblock {\em IEEE Transactions on Automatic Control}, 2024.

\bibitem{rodonyi2019heterogeneous}
G{\'a}bor R{\"o}d{\"o}nyi.
\newblock Heterogeneous string stability of unidirectionally interconnected mimo lti systems.
\newblock {\em Automatica}, 103:354--362, 2019.

\bibitem{silva2021string2}
Guilherme~Fr{\'o}es Silva, Alejandro Donaire, Aaron McFadyen, and Jason~J Ford.
\newblock String stable integral control design for vehicle platoons with disturbances.
\newblock {\em Automatica}, 127:109542, 2021.

\bibitem{vargas2025stochastic}
Francisco~J Vargas, Marco~A Gordon, Andr{\'e}s~A Peters, and Alejandro~I Maass.
\newblock On stochastic string stability with applications to platooning over additive noise channels.
\newblock {\em Automatica}, 171:111923, 2025.

\bibitem{riehl2022string}
James~R Riehl, Esteban~AL Hufstedler, Philippe Chatelain, and Julien~M Hendrickx.
\newblock String stability of energy-saving aircraft formations.
\newblock {\em Journal of Guidance, Control, and Dynamics}, 45(5):935--943, 2022.

\bibitem{kayacan2017multiobjective}
Erkan Kayacan.
\newblock Multiobjective $ h_{\infty}$ control for string stability of cooperative adaptive cruise control systems.
\newblock {\em IEEE Transactions on Intelligent Vehicles}, 2(1):52--61, 2017.

\bibitem{zhou2023data}
Yang Zhou, Xinzhi Zhong, Qian Chen, Soyoung Ahn, Jiwan Jiang, and Ghazaleh Jafarsalehi.
\newblock Data-driven analysis for disturbance amplification in car-following behavior of automated vehicles.
\newblock {\em Transportation research part B: methodological}, 174:102768, 2023.

\bibitem{dawson2023safe}
Charles Dawson, Sicun Gao, and Chuchu Fan.
\newblock Safe control with learned certificates: A survey of neural lyapunov, barrier, and contraction methods for robotics and control.
\newblock {\em IEEE Transactions on Robotics}, 39(3):1749--1767, 2023.

\bibitem{yang2023model}
Yujie Yang, Yuxuan Jiang, Yichen Liu, Jianyu Chen, and Shengbo~Eben Li.
\newblock Model-free safe reinforcement learning through neural barrier certificate.
\newblock {\em IEEE Robotics and Automation Letters}, 8(3):1295--1302, 2023.

\bibitem{zhao2020synthesizing}
Hengjun Zhao, Xia Zeng, Taolue Chen, and Zhiming Liu.
\newblock Synthesizing barrier certificates using neural networks.
\newblock In {\em Proceedings of the 23rd international conference on hybrid systems: Computation and control}, pages 1--11, 2020.

\bibitem{10886052}
Manan Tayal, Hongchao Zhang, Pushpak Jagtap, Andrew Clark, and Shishir Kolathaya.
\newblock Learning a formally verified control barrier function in stochastic environment.
\newblock In {\em 2024 IEEE 63rd Conference on Decision and Control (CDC)}, pages 4098--4104, 2024.

\bibitem{10591251}
Xinyu Wang, Luzia Knoedler, Frederik~Baymler Mathiesen, and Javier Alonso-Mora.
\newblock Simultaneous synthesis and verification of neural control barrier functions through branch-and-bound verification-in-the-loop training.
\newblock In {\em 2024 European Control Conference (ECC)}, pages 571--578, 2024.

\bibitem{zhang2023exact}
Hongchao Zhang, Junlin Wu, Yevgeniy Vorobeychik, and Andrew Clark.
\newblock Exact verification of relu neural control barrier functions.
\newblock {\em Advances in neural information processing systems}, 36:5685--5705, 2023.

\bibitem{salamati2024data}
Ali Salamati, Abolfazl Lavaei, Sadegh Soudjani, and Majid Zamani.
\newblock Data-driven verification and synthesis of stochastic systems via barrier certificates.
\newblock {\em Automatica}, 159:111323, 2024.

\bibitem{abate2024safe}
Alessandro Abate, Sergiy Bogomolov, Alec Edwards, Kostiantyn Potomkin, Sadegh Soudjani, and Paolo Zuliani.
\newblock Safe reach set computation via neural barrier certificates.
\newblock {\em IFAC-PapersOnLine}, 58(11):107--114, 2024.

\bibitem{pmlr-v270-hu25a}
Hanjiang Hu, Yujie Yang, Tianhao Wei, and Changliu Liu.
\newblock Verification of neural control barrier functions with symbolic derivative bounds propagation.
\newblock In {\em Proceedings of The 8th Conference on Robot Learning}, volume 270, pages 1797--1814. PMLR, 2024.

\bibitem{zhang2025gcbf+}
Songyuan Zhang, Oswin So, Kunal Garg, and Chuchu Fan.
\newblock Gcbf+: A neural graph control barrier function framework for distributed safe multi-agent control.
\newblock {\em IEEE Transactions on Robotics}, 2025.

\bibitem{berger2023counterexample}
Guillaume~O Berger and Sriram Sankaranarayanan.
\newblock Counterexample-guided computation of polyhedral lyapunov functions for piecewise linear systems.
\newblock {\em Automatica}, 155:111165, 2023.

\bibitem{debauche2024stability}
Virginie Debauche, Alec Edwards, Rapha{\"e}l~M Jungers, and Alessandro Abate.
\newblock Stability analysis of switched linear systems with neural lyapunov functions.
\newblock In {\em Proceedings of the AAAI Conference on Artificial Intelligence}, volume~38, pages 21010--21018, 2024.

\bibitem{liu2025physics}
Jun Liu, Yiming Meng, Maxwell Fitzsimmons, and Ruikun Zhou.
\newblock Physics-informed neural network lyapunov functions: Pde characterization, learning, and verification.
\newblock {\em Automatica}, 175:112193, 2025.

\bibitem{mueller2023certified}
Mark~Niklas Mueller, Franziska Eckert, Marc Fischer, and Martin Vechev.
\newblock Certified training: Small boxes are all you need.
\newblock In {\em The Eleventh International Conference on Learning Representations}, 2023.

\bibitem{tramer2020adaptive}
Florian Tramer, Nicholas Carlini, Wieland Brendel, and Aleksander Madry.
\newblock On adaptive attacks to adversarial example defenses.
\newblock {\em Advances in neural information processing systems}, 33:1633--1645, 2020.

\bibitem{wang2021beta}
Shiqi Wang, Huan Zhang, Kaidi Xu, Xue Lin, Suman Jana, Cho-Jui Hsieh, and J~Zico Kolter.
\newblock Beta-crown: Efficient bound propagation with per-neuron split constraints for neural network robustness verification.
\newblock {\em Advances in neural information processing systems}, 34:29909--29921, 2021.

\bibitem{wu2024marabou}
Haoze Wu, Omri Isac, Aleksandar Zelji{\'c}, Teruhiro Tagomori, Matthew Daggitt, Wen Kokke, Idan Refaeli, Guy Amir, Kyle Julian, Shahaf Bassan, et~al.
\newblock Marabou 2.0: a versatile formal analyzer of neural networks.
\newblock In {\em International Conference on Computer Aided Verification}, pages 249--264. Springer, 2024.

\bibitem{lopez2023nnv}
Diego~Manzanas Lopez, Sung~Woo Choi, Hoang-Dung Tran, and Taylor~T Johnson.
\newblock Nnv 2.0: the neural network verification tool.
\newblock In {\em International Conference on Computer Aided Verification}, pages 397--412. Springer, 2023.

\bibitem{ding2022novel}
Mi~Ding, Kaipeng Lin, Wang Lin, and Zuohua Ding.
\newblock A novel counterexample-guided inductive synthesis framework for barrier certificate generation.
\newblock In {\em 2022 IEEE 33rd International Symposium on Software Reliability Engineering (ISSRE)}, pages 263--273. IEEE, 2022.

\bibitem{xu2024eclipse}
Yuezhu Xu and S~Sivaranjani.
\newblock Eclipse: Efficient compositional lipschitz constant estimation for deep neural networks.
\newblock {\em Advances in Neural Information Processing Systems}, 37:10414--10441, 2024.

\bibitem{nnvTwo}
Diego~Manzanas Lopez, Sung~Woo Choi, Hoang{-}Dung Tran, and Taylor~T. Johnson.
\newblock {{NNV} 2.0: The Neural Network Verification Tool}.
\newblock In Constantin Enea and Akash Lal, editors, {\em International Conference on Computer Aided Verification}, volume 13965, pages 397--412. Springer, 2023.

\bibitem{xu2020automatic}
Kaidi Xu, Zhouxing Shi, Huan Zhang, Yihan Wang, Kai-Wei Chang, Minlie Huang, Bhavya Kailkhura, Xue Lin, and Cho-Jui Hsieh.
\newblock {Automatic Perturbation Analysis for Scalable Certified Robustness and Beyond}.
\newblock {\em Advances in Neural Information Processing Systems}, 33:1129--1141, 2020.

\bibitem{verinet}
Patrick Henriksen and Alessio~R. Lomuscio.
\newblock {Efficient Neural Network Verification via Adaptive Refinement and Adversarial Search}.
\newblock In Giuseppe~De Giacomo, Alejandro Catal{\'{a}}, Bistra Dilkina, Michela Milano, Sen{\'{e}}n Barro, Alberto Bugar{\'{\i}}n, and J{\'{e}}r{\^{o}}me Lang, editors, {\em European Conference on Artificial Intelligence}, volume 325, pages 2513--2520. {IOS} Press, 2020.

\bibitem{amos2017input}
Brandon Amos, Lei Xu, and J~Zico Kolter.
\newblock Input convex neural networks.
\newblock In {\em International conference on machine learning}, pages 146--155. PMLR, 2017.

\bibitem{jiang2001full}
Rui Jiang, Qingsong Wu, and Zuojin Zhu.
\newblock Full velocity difference model for a car-following theory.
\newblock {\em Physical Review E}, 64(1):017101, 2001.

\bibitem{kellett2014compendium}
Christopher~M Kellett.
\newblock A compendium of comparison function results.
\newblock {\em Mathematics of Control, Signals, and Systems}, 26(3):339--374, 2014.

\bibitem{gallian2021contemporary}
Joseph Gallian.
\newblock {\em Contemporary abstract algebra}.
\newblock Chapman and Hall/CRC, 2021.

\bibitem{barber1996quickhull}
C~Bradford Barber, David~P Dobkin, and Hannu Huhdanpaa.
\newblock The quickhull algorithm for convex hulls.
\newblock {\em ACM Transactions on Mathematical Software (TOMS)}, 22(4):469--483, 1996.

\end{thebibliography}
\appendix
\section{Proofs}

\subsection{Proof of Theorem~\ref{thm: siss-lyap}}
\begin{proof}
Define the composite Lyapunov function:
\begin{equation}
V(k) = \max_{i\in \mathcal{N}} V_i(x_{i,k}).\label{eq: composite lyapunov}
\end{equation}
Then, taking the maximum over $i$ in the decremental condition in Eq.~\eqref{eq: local_sISS_decrement} and using $V_j(x_{j,k}) \leq V(k)$, we obtain
\begin{align}
V(k+1) &\leq \max_{i\in\mathcal{N}}   \sum_{j\in\mathcal{E}_i\cup\{i\}} \gamma_{i,j}V_j(x_{j,k})  + \psi\max_{i\in\mathcal{N}}  |d_{i,k}|_2\notag\\
&\leq \max_i\sum_{j\in\mathcal{E}_i \cup \{i\}}\gamma_{i,j}V(k)+\psi |d_k|_2 \label{eq:V_eps0}
\end{align}
where $|d_k|_2 :
= \max_{i\in\mathcal{N}}|d_{i,k}|_2$. 

By the small-gain condition, there exists \(\varepsilon > 0\) such that
\begin{equation}
\max_{i\in\mathcal{N}} \sum_{j\in\mathcal{E}_i\cup\{i\}} \gamma_{i,j} \leq (1 - \varepsilon),
\end{equation}
Substituting into Eq.~\eqref{eq:V_eps0} yields
\begin{equation}
V(k+1) \leq (1 - \varepsilon) V(k) + \psi|d_k|_2. \label{eq:V_eps}
\end{equation}
Iterating Eq.~\eqref{eq:V_eps}, we obtain 
\begin{align}
V(k) &\leq (1 - \varepsilon)^k V(0) + \sum_{m=0}^{k-1} (1 - \varepsilon)^{k-1-m} \psi |d_m|_2 \notag \\
& \leq (1 - \varepsilon)^k V(0) + \frac{1 - (1 - \varepsilon)^k}{\varepsilon} \psi\max_{m < k}\left( |d_m|_2 \right) \notag \\
& \leq (1 - \varepsilon)^k V(0) + \frac{1}{\varepsilon} \psi \max_{m < k}|d_m|_2 
\label{eq: Gronwall}
\end{align}
Since $\alpha_1\in \mathcal{K}_\infty$ is strictly increasing in Eq.~\eqref{eq: local_sISS_bound}, we obtain
\begin{subequations}
\begin{align}
&|x_{i,k}|_2\le \alpha_1^{-1}(V_i(x_{i,k})) \le \alpha_1^{-1}(V(k))\label{eq: prove_dsiss_2}\\
&\le \alpha_1^{-1}\left((1-\varepsilon)^k V(0)  
+ \frac{1}{\varepsilon} \psi\max_{m < k} |d_m|_2 \right)\label{eq: prove_dsiss_3}
\end{align}
\end{subequations}
where Eq.~\eqref{eq: prove_dsiss_2} directly follows Eqs.~\eqref{eq: local_sISS_bound} and \eqref{eq: composite lyapunov}, and Eq.~\eqref{eq: prove_dsiss_3} from Eq.~\eqref{eq: Gronwall}. 

In Eq.~\eqref{eq: prove_dsiss_3}, $V(0)$ can be upper-bounded using Eq.~\eqref{eq: local_sISS_bound},
\begin{equation}
V(0) = \max_{j\in\mathcal{N}} V_j(x_{j,0})
\le \alpha_2\!\left(\max_{j\in\mathcal{N}} |x_{j,0}|_2\right).
\label{eq: bound_on_initial}
\end{equation}
The term  $\sup_{m<k}|d_m|_2$ can be bounded by 
\begin{align}
\sup_{m<k}|d_{j,m}|_2
&= \max_{m<k}\max_{j\in\mathcal N}|d_{j,m}|_2 \notag\\
&= \max_{j\in\mathcal N}\max_{m<k}|d_{j,m}|_2 \notag\\
&\le \max_{j\in\mathcal N}\sup_{m\in\mathbb N}|d_{j,m}|_2\notag\\
&= \max_{j\in\mathcal N}\|d_j\|_{\mathcal L_\infty}. \label{eq:V_eps_2}
\end{align}
Substituting Eqs.~\eqref{eq: bound_on_initial} and \eqref{eq:V_eps_2} into Eq.~\eqref{eq: prove_dsiss_3}, and applying the triangle inequality in the form $\alpha_1^{-1}(a+b)\le\alpha_1^{-1}(2a)+\alpha_1^{-1}(2b)$ (see \cite[Lemma~9]{kellett2014compendium}), we obtain
\begin{align}
|x_{i,k}|_2
\le &
\alpha_1^{-1}\!\left(
    2(1-\varepsilon)^k\,\alpha_2\!\left(\max_{j\in\mathcal{N}}|x_{j,0}|_2\right)
     \right)+\notag\\
&\alpha_1^{-1}\left(\frac{2}{\varepsilon}\psi\max_{j\in\mathcal N}\!\big(\|d_j\|_{\mathcal L_\infty}\big)
\right)\label{eq: prove_dsiss_5}
\end{align}
This yields the $\mathcal{KL}$ and $\mathcal{K}_\infty$ functions in Definition~\ref{def: siss} as
\begin{align}
\beta(s,k) := \alpha_1^{-1}\!\big(2(1-\varepsilon)^k \alpha_2(s)\big), 
\quad
\mu(r) := \alpha_1^{-1}\!\left(\tfrac{2}{\varepsilon} \psi r\right), \notag 
\end{align}
which are independent of the network size $N$. This completes the proof.
\end{proof}

\subsection{Proof of Theorem~\ref{thm: verify_neural_network}}
\begin{proof}
To simplify notation, denote $z = (x_i, \{x_j\}_{j \in \mathcal{E}_i}, d_i) \in \mathcal{Z}_i$. Then the representations of true and approximated dynamics can be simplified as 
\begin{align}
&f_i(z) = f_i(x_i, \{x_j\}_{j \in \mathcal{E}_i}, d_i),\\
&\tilde{f}_i(z) = \tilde{f}_i(x_i, \{x_j\}_{j \in \mathcal{E}_i}, d_i),
\end{align}
which are Lipschitz continuous with constants $L_{f_i}$ and $L_{\tilde{f}_i}$, respectively, by Assumption~\ref{asmp: Lipschitz}. Then, the error function $e_i(z) = f_i(z) - \tilde{f}_i(z)$ is also Lipschitz continuous with constant $L_{e_i} \leq (L_{f_i} + L_{\tilde{f}_i})$.

We seek to bound the error function $|e_i(z)|_2$ for any  $z\in\mathcal{R}_i$. For $z'\in \mathcal{D}_i$ being the nearest grid point to $z$, we have
\begin{align}
|e_i(z)|_2 &= |e_i(z) - e_i(z') + e_i(z')|_2 \notag\\
&\leq |e_i(z) - e_i(z')|_2 + |e_i(z')|_2 \notag \\
&\leq (L_{f_i} + L_{\tilde{f}_i}) |z - z'|_2 + |e_i(z')|_2
\end{align}
Here, note that $\mathcal{D}_i\subset \mathcal{Z}_i$ is a uniform grid with per-dimension step sizes $\boldsymbol{\Delta}$. By the condition of this theorem, the maximum Euclidean distance between any $z \in \mathcal{Z}_i$ and its nearest grid point $z' \in \mathcal{D}_i$ is $r = \frac{1}{2} |\boldsymbol{\Delta}|_2$. Then, 
\begin{align}
|e_i(z)|_2 &\leq (L_{f_i} + L_{\tilde{f}_i}) |z - z'|_2 + |e_i(z')|_2 \notag \\
&\leq \frac{1}{2} (L_{f_i} + L_{\tilde{f}_i}) |\boldsymbol{\Delta}|_2 + \hat{\epsilon}_i = \epsilon_i
\end{align}
where the last inequality uses Assumption~\ref{asmp: uniform sample}, i.e., $|e_i(z')|_2 \leq \hat{\epsilon}_i$ for any $z' \in \mathcal{D}_i$.

Consequently, for any $z_k = (x_{i,k}, \{x_{j,k}\}_{j \in \mathcal{E}_i}, d_{i,k}) \in \mathcal{Z}_i$, the next states satisfy $|x_{i,k+1} - \tilde{x}_{i,k+1}|_2 = |e_i(z_k)|_2 \leq \epsilon_i$. Then, by the Lipschitz continuity of $V_i$,
\begin{align}
&V_i(x_{i,k+1}) \leq V_i(\tilde{x}_{i,k+1}) + L_{V_i} \epsilon_i \notag \\
\leq &  \sum_{j \in \mathcal{E}_i\cup\{i\}} \gamma_{i,j}V_j({x}_{j,k}) + \psi|d_{i,k}|_2- \delta_i + L_{V_i}\epsilon_i \label{eq:thm2_main1}\\
\leq & \sum_{j \in \mathcal{E}_i\cup\{i\}} \gamma_{i,j}V_j({x}_{j,k}) + \psi|d_{i,k}|_2  \label{eq:thm2_main2}
\end{align}
where Eq.~\eqref{eq:thm2_main1} is derived from Eq.~\eqref{eq: new_siss_decremental}, and Eq.~\eqref{eq:thm2_main2} from $\delta_i \geq L_{V_i}\epsilon_i$. 

Hence, the decremental condition holds for the true system. Together with the small-gain condition Eq.~\eqref{eq: sISS_small_gain} and class-$\mathcal{K}$ bounds Eq.~\eqref{eq: local_sISS_bound}, this implies the true system satisfies all conditions in Theorem~\ref{thm: siss-lyap} and hence is discrete-time sISS. This completes the proof.
\end{proof}

\subsection{Proof of Theorem~\ref{thm:Certifiable Equivalence}}
\begin{proof}
We aim to show that for any $\{\tilde{x}_{i,k}\}_{i \in \widetilde{\mathcal{N}}} \in  \prod_{i \in \widetilde{\mathcal{N}}}\widetilde{\mathcal{R}}_{i}$ and $\{\tilde{d}_{i,k}\}_{i \in \widetilde{\mathcal{N}}}$, the conditions of Theorem~\ref{thm: siss-lyap} hold. 

Since the interconnected system $\widetilde{\mathcal{I}}$ is substructure-isomorphic to $\mathcal{I}$, there exists an injective mapping $\tau: \widetilde{\mathcal{N}} \rightarrow \mathcal{N}$ such that $\widetilde{\mathcal{E}}_i = \mathcal{E}_{\tau(i)}$ and $\widetilde{f}_i = f_{\tau(i)}$ for each $i \in \widetilde{\mathcal{N}}$. Then, it follows that the domain of $\widetilde{f}_i$, namely $\widetilde{\mathcal{R}}_{i}$, satisfies $\widetilde{\mathcal{R}}_{i}=\mathcal{R}_{\tau(i)}$. Hence, we can define $\{x_{i,k}\}_{i \in \mathcal{N}} \in  \prod_{i \in \mathcal{N}}\mathcal{R}_{i}$ and $\{d_{i,k}\}_{i \in \widetilde{\mathcal{N}}}$ satisfying $\tilde{x}_{i,k} = x_{\tau(i),k},\tilde{d}_{i,k}=d_{\tau(i),k},~\forall i \in \widetilde{\mathcal{N}}$.  
Then, it follows that 
\begin{align}
    \widetilde{x}_{i,k+1} &= \widetilde{f}_{i}(\widetilde{x}_{i,k}, \{\widetilde{x}_{j,k}\}_{j\in\widetilde{\mathcal{E}}_i},\widetilde{d}_{i,k}) \notag\\
&= f_{\tau(i)}(x_{\tau(i),k}, \{x_{j,k}\}_{j\in\tau(\mathcal{E}_i)},d_{\tau(i),k})\notag\\
&= x_{\tau(i),k+1}  \label{eq:thm3_x_k+1}
\end{align}
As system $\mathcal{I}$ satisfies the conditions of Theorem~\ref{thm: siss-lyap}, we have 

\noindent 1. There exists $\alpha_1,\alpha_2 \in \mathcal{K}_\infty$ such that $\forall i \in \widetilde{\mathcal{N}}$
\begin{equation}
\alpha_1(|x_{\tau(i),k}|_2) \leq V_{\tau(i)}(x_{\tau(i),k}) \leq \alpha_2(|x_{\tau(i),k}|_2),
\end{equation}
\noindent 2. There exists positive gains \(\Gamma\)  satisfying the small-gain condition
such that the decremental inequality holds 
\begin{align}
   & V_{\tau(i)}\bigl(x_{\tau(i),k+1}\bigr) - \gamma_{\tau(i),\tau(i)} V_{\tau(i)}(x_{\tau(i),k}) \notag\\
&-  \sum_{j \in \mathcal{E}_{i}} \gamma_{\tau(i),\tau(j)} V_{\tau(j)}(x_{\tau(j),k}) - \psi |d_{\tau(i),k}|_2 \le 0
\end{align}
Define gains for system $\widetilde{\mathcal{I}}$ as $\widetilde{\gamma}_{i,j} := \gamma_{\tau(i),\tau(j)}$ and certificates $\{\widetilde{V}_i\}_{i\in\mathcal{\widetilde{N}}}$ as $\widetilde{V}_i := V_{\tau(i)}, \forall i \in \widetilde{\mathcal{N}}$. Then, the small-gain condition is satisfied, and by Eq.~\eqref{eq:thm3_x_k+1} we have $\widetilde{V}_i(\widetilde{x}_{i,k}) = V_{\tau(i)}(x_{\tau(i),k})$ and $\widetilde{V}_i(\widetilde{x}_{i,k+1}) = V_{\tau(i)}(x_{\tau(i),k+1})$.

Therefore, we can calculate
\begin{align}
& \alpha_1(|\tilde{x}_{i,k}|_2) \leq \widetilde{V}_{i}(\tilde{x}_{i,k}) \leq \alpha_2(|\tilde{x}_{i,k}|_2) \\
&\widetilde{V}_i\bigl(\widetilde{x}_{i,k+1}\bigr) \leq \sum_{j \in \widetilde{\mathcal{E}}_i\cup\{i\}} \widetilde{\gamma}_{i,j} \widetilde{V}_j(\widetilde{x}_{j,k}) - \psi |\widetilde{d}_{i,k}|_2  
\end{align}
Thus, the conditions in Theorem~\ref{thm: siss-lyap} hold for system $\widetilde{\mathcal{I}}$. This completes the proof. 
\end{proof}

\subsection{Proof of Theorem~\ref{thm:eq_nodes}}
\begin{proof}
By Def.~\ref{dfn:eq_nodes}, there exists a permutation $\tau$ such that $\tau(\mathcal{E}_{i}) = \mathcal{E}_{\tau(i)}$ and $f_{i} = f_{\tau(i)}$ for each $i\in\mathcal{N}$. By Theorem~5.3 in Gallian~\cite{gallian2021contemporary}, any permutation of a finite set has a finite order, i.e., there exists $M \in \mathbb{Z}_+$ such that $\tau^M(i)=i,~\forall i\in\mathcal{N}$. 

Clearly, the permutations $\tau^m,~m=0,\cdots,M-1$, satisfy the conditions of Theorem~\ref{thm:Certifiable Equivalence}. Then, the system admits sISS certificates $\{V_{\tau^m(i)}\}_{i \in \mathcal{N}}$ such that for any $i\in\mathcal{N}$, any $\{x_{j,k}\}_{j\in\mathcal{E}_i\cup \{i\}}\in\prod_{j\in\mathcal{E}_i\cup \{i\}}\mathcal{R}_{j}$, and any $m=0,\cdots,M-1$, the following inequalities hold: 
\begin{align}
&\alpha_1(|x_{i,k}|_2) \leq V_{\tau^m(i)}(x_{i,k}) \leq \alpha_2(|x_{i,k}|_2), \label{eq:thm3-siss1-discrete} \\
&V_{\tau^m(i)}(x_{i,k+1}) - \sum_{j \in \mathcal{E}_{i} \cup \{i\}} \gamma_{\tau^m(i),\tau^m(j)} V_{\tau^m(j)}(x_{j,k}) 
\notag\\
& - \psi|d_{i,k}|_2 \ \le \ 0, \label{eq:thm3-siss2-discrete}
\end{align}
where $x_{i,k+1}=f_{\tau(i)}(x_{i,k},\{x_{j,k}\}_{j\in\mathcal{E}_{i}},d_{i,k})\linebreak =f_{i}(x_{i,k},\{x_{j,k}\}_{j\in\mathcal{E}_{i}},d_{i,k})$ is independent of $\tau^m$.

Define the max-aggregated certificate
\begin{align}
    \widetilde{V}_{i}(x_{i,k}) := \max_{m=0,\cdots,M-1} V_{\tau^m(i)}(x_{i,k}),~\forall i\in\mathcal{N} \label{eq:thm3_tilde_V}
\end{align} 
with the active index set $\mathcal{A}_{i,k} := \{m\mid \widetilde{V}_{i}(x_{i,k})=V_{\tau^m(i)}(x_{i,k})\}$ indicating the values of $m$ at which the maximum is attained. As $\tau$ has an order of $M$, $\widetilde{V}_{i}$ is invariant under permutation $\tau$, i.e., $\widetilde{V}_{i} = \widetilde{V}_{\tau(i)}$. 

Pick $m^*\in\mathcal A_{i,k+1}$. Then by Eq.~\eqref{eq:thm3-siss1-discrete}, $\widetilde{V}_{i} = V_{\tau^{m^*}(i)}$ satisfies the class -$\mathcal{K}$ condition. Moreover, from Eq.~\eqref{eq:thm3-siss2-discrete}, 
\begin{align}
&\widetilde{V}_i(x_{i,k+1})=V_{\tau^{m^{*}}(i)}(x_{i,k+1}) \notag\\
& \le\sum_{j \in \mathcal{E}_{i}\cup\{i\}} \gamma_{\tau^{m^*}(i),\tau^{m^*}(j)} V_{\tau^{m^*}(j)}(x_{j,k}) 
+ \psi |d_{i,k}|_2 \notag \\
&\le \sum_{j \in \mathcal{E}_{i}\cup\{i\}}\gamma_{\tau^{m^*}(i),\tau^{m^*}(j)}\widetilde{V}_j(x_{j,k})
+\psi|d_{i,k}|_2
\end{align}
where the last inequality is derived from Eq.~\eqref{eq:thm3_tilde_V}, i.e., $V_{\tau^{m^*}(j)}(x_{j,k})  \le \widetilde{V}_j(x_{j,k}),~\forall j \in \mathcal{N}$. Clearly, $\widetilde{\Gamma} = \{\gamma_{\tau^{m^*}(i),\tau^{m^*}(j)}\}_{i\in\mathcal{N}, j\in\mathcal{E}_i\cup\{i\}}$ satisfies the small-gain condition. 

Therefore, all conditions in Theorem \ref{thm: siss-lyap} hold, and the system admits Lyapunov certificates $\{\widetilde{V}_i\}_{i\in\mathcal{N}}$ with $\widetilde{V}_i = \widetilde{V}_{\tau(i)},~\forall i \in \mathcal{N}$. This completes the proof. 
\end{proof}

\subsection{Proof of Theorem~\ref{thm:siss_affine_last_layer}}
\begin{proof}
Fix any $\chi \in \mathcal{B}$.  
Since $\mathcal{B} = \operatorname{conv}\{\chi^{\nu}\}_{\nu \in \Omega}$, there exist $\lambda_\nu\ge0$ with $\sum_{\nu} \lambda_\nu = 1$ such that
\begin{equation}
    \chi = \sum_{\nu \in \Omega} \lambda_\nu \chi^{\nu}.
\end{equation}
By Eq.~\eqref{eq:ca_dynamics}, the affine dependence of $f_i$ on $\chi$,
\begin{equation}
    x_{i,k+1}(\chi) = \sum_{\nu \in \Omega} \lambda_\nu\, x_{i,k+1}(\chi^{\nu}).
\end{equation}
By the convexity of $V_i$ (Jensen’s inequality),
\begin{equation}
    V_i\bigl(x_{i,k+1}(\chi)\bigr)
    \le \sum_{\nu \in \Omega} \lambda_\nu\, V_i\bigl(x_{i,k+1}(\chi^{\nu})\bigr).
\end{equation}
Multiplying Eq.~\eqref{eq:local_siss_vertex} for each $\chi^\nu$ by $\lambda_\nu$ and summing over $\nu$ yields
\begin{align}
&\sum_{\nu\in\Omega} \lambda_\nu V_i\bigl(x_{i,k+1};\chi^{\nu}\bigr) 
\leq  \sum_{j \in \mathcal{E}_i\cup\{i\}} \gamma_{i,j} V_j(x_{j,k})  + \psi |d_{i,k}|_2.
\end{align}
Replacing the first term by the Jensen bound gives exactly Eq.~\eqref{eq:local_siss_vertex} for $\chi$.  
This completes the proof.
\end{proof}

\subsection{Proof of Theorem~\ref{thm:augmentedSystem}}
\begin{proof}
As subsystem $\mathcal{I}' = (\mathcal{N}', \{\mathcal{E}'_i\}_{i\in\mathcal{N}'}, \{f_i'\}_{i\in\mathcal{N}'})$ is already sISS-verified under certificates $\{V_i'\}_{i \in \mathcal{N}'}$ satisfying Theorem~\ref{thm: siss-lyap}, then for any $i\in\mathcal{N}'$ and $\{x_{j}\}_{j\in\mathcal{E}_i'}\in\prod_{j\in\mathcal{E}'_i}\mathbb{R}^{n_j}$, we have:
\begin{align}
&\alpha_1(|x_{i,k}|_2)\le V_i'(x_{i,k}) \le \alpha_2(|x_{i,k}|_2), \label{eq:orig_siss_1_discrete} \\
&V_i'\bigl(x_{i,k+1}\bigr) \le  \sum_{j \in \mathcal{E}_i'\cup\{i\}} \gamma_{i,j} V_j'(x_{j,k}) + \psi |d_{i,k}|_2.
\label{eq:orig_siss_2_discrete}
\end{align}
Similarly, suppose the remaining agents in $\mathcal{N}''=\mathcal{N}\backslash\mathcal{N}'$ with neighbor sets $\mathcal{E}''_i\subset\mathcal{N},~i\in\mathcal{N}''$ also satisfy the sISS conditions, i.e., there exist certificates $\{V_i''\}_{i \in \mathcal{N}''}$ such that for all $i\in\mathcal{N}''$,
\begin{align}
&\alpha_1(|x_{i,k}|_2)\le V_i''(x_{i,k}) \le \alpha_2(|x_{i,k}|_2), \label{eq:new_siss_1_discrete} \\
&V_i''\bigl(x_{i,k+1}\bigr) \le \sum_{j \in \mathcal{E}_i''\cup\{i\}} \gamma_{i,j} V_j''(x_{j,k}) + \psi |d_{i,k}|_2.
\label{eq:new_siss_2_discrete}
\end{align}
Now, for the original system $\mathcal{I} = (\mathcal{N}, \{\mathcal{E}_i\}_{i\in\mathcal{N}}, \{f_i\}_{i\in\mathcal{N}})$, define controllers and certificates as
\begin{equation}
V_i =
\begin{cases}
V_i'', & i\in \mathcal{N}'' ,\\
V_i', & i\in \mathcal{N}'.
\end{cases}\label{eq:thm7_lyap}
\end{equation}
By Eqs.~\eqref{eq:new_siss_1_discrete}--\eqref{eq:new_siss_2_discrete}, we only need to verify the sISS conditions for nodes in $\mathcal{N}'$.  
Since the subsystem dynamics are independent of the remaining agents, i.e.,
$f_i(x_{i,k},\{x_{j,k}\}_{j\in \mathcal{E}_i},d_{i,k})
= f_i'(x_{i,k},\{x_{j,k}\}_{j\in \mathcal{E}_i'},d_{i,k}), \forall i \in \mathcal{N}'$, we have:
\begin{align}
&V_i\bigl(x_{i,k+1}\bigr)  - \sum_{j \in \mathcal{E}_i\cup\{i\}} \gamma_{i,j} V_j(x_{j,k})- \psi |d_{i,k}|_2 \notag \\
&=~V_i'\bigl(x_{i,k+1}\bigr) - \sum_{j \in \mathcal{E}_i'\cup\{i\}} \gamma_{i,j} V_j'(x_{j,k}) - \psi |d_{i,k}|_2 \notag \\
&\le 0,
\end{align}
where the last inequality follows from Eq.~\eqref{eq:orig_siss_2_discrete} and the fact that $\mathcal{E}_i' \subseteq \mathcal{E}_i$. Moreover, the class-$\mathcal K_\infty$ bounds in Eq.~\eqref{eq:orig_siss_1_discrete} for $V_i'$ on $\mathcal N'$ carry over to $V_i$ via the definition in Eq.~\eqref{eq:thm7_lyap}. 
Thus, the sISS conditions hold for all $i\in\mathcal{N}$, and $\mathcal{I}$ is sISS decomposable.
\end{proof}

\subsection{Proof of Corollary~\ref{col: verify_neural_network}}
\begin{proof}
To simplify notation, let $z = (x_i, \{x_j\}_{j \in \mathcal{E}_i}, d_i)$, $w = (x_i, \{x_j\}_{j\in\mathcal{E}_j})$. Let the closed-loop dynamics for the true and approximated systems be denoted by 
\begin{align}
&g_i(z) = f_i(x_i, \{x_j\}_{j \in \mathcal{E}_i}, \pi_i(x_i, \{x_j\}_{j \in \mathcal{E}_i}), d_i),\\
&\tilde{g}_i(z) = \tilde{f}_i(x_i, \{x_j\}_{j \in \mathcal{E}_i}, \pi_i(x_i, \{x_j\}_{j \in \mathcal{E}_i}), d_i),
\end{align}
We aim to calculate the Lipschitz constant of $g_i(z)$.
Under Assumption~\ref{asmp: Lipschitz control}, we have: 
\begin{align}
&|g_i(z) - g_i(z')|_2 \notag\\
&\leq L_{f_i} |(x_i - x_i', \{x_j - x_j'\}_{j\in\mathcal{E}_j}, u_i - u_i', d_i - d_i')|_2\notag \\
&\leq L_{f_i} \left( |(x_i - x_i', \{x_j - x_j'\}_{j\in\mathcal{E}_j}, d_i - d_i')|_2 + |u_i - u_i'|_2 \right)\notag \\
&= L_{f_i} \left( |z - z'|_2 + |\pi_i(w) - \pi_i(w')|_2 \right)\notag \\
&\leq L_{f_i} \left( |z - z'|_2 + L_{\pi_i} |w - w'|_2 \right) \notag\\
&\leq L_{f_i} \left( |z - z'|_2 + L_{\pi_i} |z - z'|_2 \right) \notag\\
&\leq L_{f_i} (1 + L_{\pi_i}) |z - z'|_2.
\end{align}
Thus, the Lipschitz constant of $g_i$ satisfies 
\begin{align}
L_{g_i} \leq L_{f_i}(1 + L_{\pi_i}) \label{eq: lip_g_i}
\end{align}
Similarly, the Lipschitz constant of $\tilde{g}_i$ satisfies:
\begin{align}
L_{\tilde{g}_i} \leq L_{\tilde{f}_i}(1 + L_{\pi_i}).
\label{eq: lip_tilde_g_i}
\end{align} 
The rest of the proof directly follows from the proof of Theorem~\ref{thm: verify_neural_network}.
\end{proof}

\section{Illustrative Examples}
\subsection{Illustrative Example for Definition~\ref{dfn:eq_nodes}}
\label{appendix: example 2}
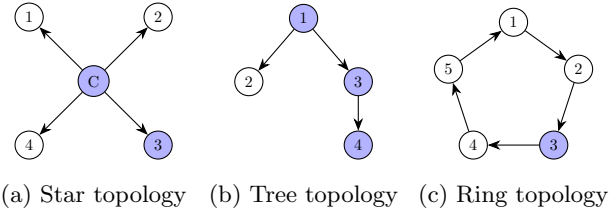
\begin{figure}[h!]
\centering
\begin{subfigure}[b]{0.15\textwidth}
\centering
\begin{tikzpicture}[->, >=Stealth, node distance=1cm, scale=0.6, transform shape]
    \tikzstyle{normal}=[draw, circle, fill=white, minimum size=6mm]
    \tikzstyle{highlight}=[draw, circle, fill=blue!30, minimum size=6mm]
    \node[highlight] (C) at (0,0) {C}; 
    \node[normal] (L2) at (45:2) {2};
    \node[highlight] (L3) at (315:2) {3};
    \node[normal] (L5) at (225:2) {4};
    \node[normal] (L6) at (135:2) {1};
    \draw[->] (C) -- (L2);
    \draw[->] (C) -- (L3);
    \draw[->] (C) -- (L5);
    \draw[->] (C) -- (L6);
\end{tikzpicture}
\caption{Star topology}
\label{subfig:star-topology}
\end{subfigure}
\begin{subfigure}[b]{0.15\textwidth}
\centering
\begin{tikzpicture}[->, >=Stealth, node distance=1.2cm, scale=0.6, transform shape]
    \tikzstyle{normal}=[draw, circle, fill=white, minimum size=6mm]
    \tikzstyle{highlight}=[draw, circle, fill=blue!30, minimum size=6mm]
    \node[highlight] (N1) at (0,0) {1};
    \node[normal] (N2) at (-1.2,-1.4) {2};
    \node[highlight] (N3) at (1.2,-1.4) {3};
    \node[highlight] (N4) at (1.2,-2.8) {4};  
    \draw[->] (N1) -- (N2);
    \draw[->] (N1) -- (N3);
    \draw[->] (N3) -- (N4);
\end{tikzpicture}
\caption{Tree topology}
\label{subfig:tree-topology}
\end{subfigure}
\begin{subfigure}[b]{0.15\textwidth}
\centering
\begin{tikzpicture}[->, >=Stealth, scale=0.6, transform shape]
    \tikzstyle{normal}=[draw, circle, fill=white, minimum size=6mm]
    \tikzstyle{highlight}=[draw, circle, fill=blue!30, minimum size=6mm]
    \node[normal] (R1) at (90:1.5) {1};
    \node[normal] (R2) at (18:1.5) {2};
    \node[highlight] (R3) at (-54:1.5) {3};
    \node[normal] (R4) at (-126:1.5) {4};
    \node[normal] (R5) at (162:1.5) {5};
    \draw[->] (R1) -- (R2);
    \draw[->] (R2) -- (R3);
    \draw[->] (R3) -- (R4);
    \draw[->] (R4) -- (R5);
    \draw[->] (R5) -- (R1);
\end{tikzpicture}
\caption{Ring topology}
\label{subfig:ring-topology}
\end{subfigure}
\caption{Representative network topologies for illustrating node structure equivalence:
\protect\subref{subfig:star-topology} a star network with one hub (C) and leaves (node~C and 3 are highlighted for sISS verification);
\protect\subref{subfig:tree-topology} a small tree (node~1,3,4 is highlighted for local verification);
\protect\subref{subfig:ring-topology} a 5-node ring (node~3 is highlighted).}
\label{fig:reducible-topology-example}
\end{figure}

Consider a star network with one central node (hub) and multiple leaf nodes. All leaves share an identical connectivity pattern (each leaf connects only to the hub) and possess identical dynamics. By verifying sISS for just one leaf (together with the hub), we ensure that all other leaves inherit the same stability property, and thus the entire star remains sISS. Figure~\ref{fig:reducible-topology-example} demonstrates this concept in the left subfigure, alongside two additional topologies (a small tree in the middle and a ring on the right) for completeness.

\subsection{Illustrative Example for Definition~\ref{dfn:additiveness}}
\label{appendix: example 1}
\begin{figure}[h]
\centering
\begin{tikzpicture}[->, >=Stealth, node distance=1.3cm]
    \tikzstyle{normal}=[draw, circle, fill=white, minimum size=6mm]
    \tikzstyle{new}=[draw, circle, fill=green!30, minimum size=6mm]
    \node[normal] (N1) at (0,0) {1};
    \node[normal, right=of N1] (N2) {2};
    \node[normal, right=of N2] (N3) {3};
    \node[new, right=of N3] (N4) {4};
    \draw[->] (N1) -- (N2);
    \draw[->] (N2) -- (N3);
    \draw[->] (N3) -- (N4);
\end{tikzpicture}
\caption{A chain of three nodes (1--2--3) verified for sISS. Adding node~4 in green only requires verifying local sISS conditions between (3,4).}
\label{fig:addable-topology-example}
\end{figure}
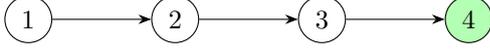

If we have a chain of $N$ nodes with sISS and we add node $N+1$ as a successor to node $N$, we need only verify local conditions for the pair $(N, N+1)$. If that local verification for node $N$ passes, the entire chain remains sISS by Theorem~\ref{thm:augmentedSystem}. See Figure~\ref{fig:addable-topology-example}.

\section{Implementation Algorithms}

\subsection{Algorithm for Theorem~\ref{thm:siss_affine_last_layer}}
\label{appendix: Algorithm for Theorem siss_affine}
\begin{algorithm}[htp]
\caption{Vertex-Hull-Based Certified Region Construction}
\label{alg:lvhv}
\begin{algorithmic}[1]
\Require Datasets $\{\mathcal D_r\}_{r\in R}$; shared embeddings $\Phi_h$; verifier $\textsc{Verify}(\Phi_h,\chi)$ for inequality~\eqref{eq: verification for all agents}
\Ensure Verified polytope $\mathcal B$ and its H-representation $(H,h)$ such that $\mathcal B=\{\chi\mid H\chi\le h\}$.
\State \textbf{Learn.} Fit the last layer for each domain:
$\chi_r \gets \arg\min_{\chi}\mathcal{L}(\chi;\mathcal D_r)$ and collect $S=\{\chi_r\}_{r\in R}$.
\State \textbf{Hull vertices.} Compute the convex hull of $S$ (e.g., Quickhull~\cite{barber1996quickhull}) and extract its convex vertex set
$\mathcal V \gets \textsc{Vertices}(\operatorname{conv}(S))$.
\State \textbf{Verify vertices.} Initialize $\mathcal V_{\mathrm{ok}}\gets\emptyset$.
\For{$\chi^\nu \in \mathcal V$}
    \If{$\textsc{Verify}(\Phi_h,\chi^\nu)$ succeeds for all agents}
        \State $\mathcal V_{\mathrm{ok}} \gets \mathcal V_{\mathrm{ok}} \cup \{\chi^\nu\}$
    \EndIf
\EndFor
\State \textbf{Polytope.} Set $\mathcal B\gets \operatorname{conv}(\mathcal V_{\mathrm{ok}})$ and compute $(H,h)$ such that
$\mathcal B=\{\chi\mid H\chi\le h\}$.
\State \textbf{Online check.} For a new parameter $\chi_{\mathrm{new}}$: if $H\chi_{\mathrm{new}}\le h$, accept without verification; otherwise call $\textsc{Verify}(\Phi_h,\chi_{\mathrm{new}})$ and, if successful, update $\mathcal V_{\mathrm{ok}}$ and recompute the $(H,h)$ of $\mathcal{B}$.
\end{algorithmic}
\end{algorithm}
Leveraging Theorem \ref{thm:siss_affine_last_layer}, the algorithm certifies a polytope of sISS controllers via vertex checking, then converts it to an $H$-representation for fast online checking, illustrating its practical utility.

\subsection{Algorithm for efficient synthesis and verification}
\label{appendix:Algorithm for Constructing Minimum Verification Network}
The algorithm constructs a Minimum Verification Network $\mathcal I^*$ by pruning a large interconnected system via the proposed Theorems. (i) \emph{Substructure isomorphism} (Def.~\ref{dfn:stru_eq}, Thm.~\ref{thm:Certifiable Equivalence}): subgraphs isomorphic to previously verified templates are identified and excluded from further verification. (ii) \emph{Node structural equivalence} (Def.~\ref{dfn:eq_nodes}, Thm.~\ref{thm:eq_nodes}): the remaining nodes are partitioned into equivalence classes, and only one representative per class is kept for verification. (iii) \emph{Modular decomposability} (Def.~\ref{dfn:additiveness}, Thm.~\ref{thm:augmentedSystem}): already verified subsystems whose dynamics are independent of external states are removed. The resulting $\mathcal I^*$ contains only class representatives and unverified modules.
\begin{algorithm}[htp]
\caption{Minimum Verification Network}
\label{alg:mvn-reduction}
\begin{algorithmic}[1]
\Require Interconnected system $\mathcal I=(\mathcal N,\{\mathcal E_i\}_{i\in\mathcal N},\{f_i\}_{i\in\mathcal N})$; optional library $\mathbb L$ of verified substructures
\Ensure Reduced network $\mathcal I^*$ 
\State \textbf{Init} $\mathcal N_{\mathrm{todo}}\gets\mathcal N$
\State \textbf{Substructure reuse (Thm.~\ref{thm:Certifiable Equivalence}).}
\For{each $(\widehat{\mathcal I},\widehat{\bm V})\in\mathbb L$}
  \For{each injective $\tau:\widehat{\mathcal N}\!\to\!\mathcal N_{\mathrm{todo}}$ with $\tau(\widehat{\mathcal E}_j)=\mathcal E_{\tau(j)}$ and $\widehat f_j=f_{\tau(j)}$}
    \State $\mathcal N_{\mathrm{todo}}\gets \mathcal N_{\mathrm{todo}}\setminus \tau(\widehat{\mathcal N})$ 
  \EndFor
\EndFor
\State \textbf{Node equivalence (Thm.~\ref{thm:eq_nodes}).}
Let $\{C_r\}_{r=1}^{R}$ be the distinct equivalence classes.
Choose one representative $i_r\in C_r$ for each $r\in \{1,\cdots,R\}$ and set
$\mathcal N_{\mathrm{todo}}\gets\{i_r \mid r\in \{1,\cdots,R\}\}$.
\State \textbf{Modular decomposition (Thm.~\ref{thm:augmentedSystem}).}
If there exists an already-verified subsystem
$\mathcal I'=(\mathcal N',\{\mathcal E'_i\}_{i\in\mathcal{N}'},\{f'_i\}_{i\in\mathcal{N}'})\subseteq\mathcal I$
such that, for every $i\in\mathcal N'$, $\mathcal E_i\subseteq\mathcal N'$
(no dependence on external states) and $f_i(\cdot)\equiv f'_i(\cdot)$ on the
restricted domain, then set
$\mathcal N_{\mathrm{todo}}\gets \mathcal N_{\mathrm{todo}}\setminus \mathcal N'$.
\State \textbf{Construct $\mathcal I^*$.} Let $S\gets\mathcal N_{\mathrm{todo}}$ and set
$\displaystyle
\mathcal I^* =
\Big(
S,\{\mathcal{E}_i\}_{i\in S},\{f_i\}_{i\in S}
\Big).
$
\State \Return $\mathcal I^*$
\end{algorithmic}
\end{algorithm}

\end{document}